\documentclass{aa}  

\usepackage{physics}
\usepackage{graphicx}
\usepackage{placeins}

\usepackage{tikz}
\definecolor{lime}{HTML}{A6CE39}
\DeclareRobustCommand{\orcidicon}{%
  \begin{tikzpicture}
  \draw[lime, fill=lime] (0,0) 
  circle [radius=0.16] 
  node[white] {{\fontfamily{qag}\selectfont \tiny ID}};
  \draw[white, fill=white] (-0.0625,0.095) 
  circle [radius=0.007];
  \end{tikzpicture}
  \hspace{-2mm}
}

\foreach \x in {A, ..., Z}{%
  \expandafter\xdef\csname orcid\x\endcsname{\noexpand\href{https://orcid.org/\csname orcidauthor\x\endcsname}{\noexpand\orcidicon}}
}

\usepackage{xcolor}
\newcommand\REV[1]{\textcolor{black}{#1}}
\usepackage{txfonts}
\begin{document}

	\title{Dynamics of the galactic component of Abell S1063 and MACS J1206.2$-$0847}
  \titlerunning{Dynamics of the galactic component of AS1063 and M1206}
  \authorrunning{G. Ferrami et al.}

    \author{G. Ferrami\inst{1,2\thanks{current affiliation, \email{gferrami@student.unimelb.edu.au} }} \orcidA{}
			\and
			G. Bertin \inst{1}  \orcidB{}
			\and
			C. Grillo \inst{1}\fnmsep\inst{3}  \orcidC{}
      \and
      A. Mercurio \inst{4}\fnmsep\inst{5}  \orcidD{}
      \and
      P. Rosati \inst{6} \orcidE{}
    }

  	\institute
      {Dipartimento di Fisica, Università degli Studi di Milano, via Celoria 16, I-20133 Milano, Italy;
		\and
      School of Physics, University of Melbourne, Parkville, VIC 3010, Australia;
		\and
			INAF—IASF Milano, via A. Corti 12, I-20133 Milano, Italy;
    \and
      Dipartimento di Fisica “E.R. Caianiello”, Università degli Studi di Salerno, Via Giovanni Paolo II 132, I-84084 Fisciano (SA), Italy;
    \and
      INAF-Osservatorio Astronomico di Capodimonte, Via Moiariello 16, 80131 Napoli, Italy;
    \and
      Dipartimento di Fisica e Scienze della Terra, Università degli Studi di Ferrara, via Saragat 1, I-44122 Ferrara, Italy
    }

   \date{Received <month day, year>; accepted <month day, year>}

 
  \abstract
  {The galactic component in clusters is commonly thought to be generally nonrotating and in a dynamical state different from that of a collisionally relaxed system. In practice, a test of such a picture is often not available.}
  {We consider the member galaxies of two clusters, Abell S1063 and MACS J1206.2$-$0847, and study the possible presence of mean rotation and some properties of their distribution in phase space. We look for empirical evidence of factors normally found in collisionally relaxed systems and others characteristic of violently-relaxed collisionless systems.}
  {Starting from the CLASH-VLT data, we obtain positions, stellar masses, and individual line-of-sight velocities for a large number of galaxies $(N_\text{AS1063} \approx 1200$ and $N_\text{M1206} \approx 650)$ extending out to $\approx 1.6$ (Abell) and $\approx 2.5$ (MACS) times the radius $r_{200}$.
  We study the spatial distribution of the galaxy velocities and the properties of the available galaxy sets when divided in stellar mass bins.
  To test the presence of velocity dispersion anisotropy we compare the results based on the Jeans equations with those obtained by assuming a specific form of the galaxy distribution function incorporating the picture of violent relaxation, where the total gravitational potential is imposed as set by the available gravitational lensing observations.}
  {We find evidence of systematic rotation in both clusters, with significant rotation in each core (within $0.5'$ from the center) and no signatures of rotation at large radii.
  While no signs are found of energy equipartition, there is a clear indication of (stellar) mass segregation. 
  Velocity dispersion anisotropy is present and qualitatively similar to that found in violently relaxed collisionless systems; 
  this last conclusion is strengthened by the overall success in matching the observations with the predictions of the physically justified distribution function.}
   {}


   \maketitle
%

\section{Introduction}

Clusters of galaxies are the largest gravitationally bound structures in the universe. Their total mass is made of three main components: a dark matter (DM) halo, a hot and optically thin plasma (the Intra-Cluster Medium, or ICM), and hundreds or thousands of galaxies (with their baryonic and dark constituents). The temperature of the ICM is consistent with the observed line-of-sight velocities of galaxies and suggests that the two visible components are in a state of quasi-equilibrium within a common gravitational potential well. The mass of galaxies and hot plasma is not sufficient to explain the depth of the potential well, which implies that most of the mass in clusters is in the form of DM (\citealt{morph_classification}, \citealt{Kravtsov_Borgani_review_2012}).
Apparently, the luminous stellar component is just a tracer of the total gravitational field in clusters, because up to 85\% of the total mass of a cluster is in the form of dark matter, whereas the mass of the Intra-Cluster Medium is generally thought to exceed that of the galactic component, as shown in studies that combine gravitational lensing and hydrostatic
ICM models (e.g., see  \citealt{Bonamigo_alignment}, \citealt{Sartoris_R2248}).\\
\indent Of course, the interplay between shapes, angular momentum, and velocity dispersion anisotropy \REV{(also referred to as pressure anisotropy)} of the three main components of clusters can provide important clues about the interpretation of specific cases but can also be a source of enormous complexity. Therefore, it would be very important to have a clear empirical determination of the dynamical state of the visible components, that is, galaxies and ICM. Examples that may give an idea of how vast is the general research area related to these concepts are the following. In cosmology, we would like to clarify the role of tidal torques on the angular momentum of structures formed in the course of evolution, through theoretical arguments (see \citealt{Peebles_origin_ang_momentum}), simulations (e.g., see \citealt{spin_parameter_TTT}) or observations (e.g., see \citealt{Filament_rotation}). In the galactic context, under the assumption that the hot X-ray emitting plasma is in hydrostatic equilibrium and traces the underlying potential well, great expectations were placed in interpreting the observed misalignment between X-ray isophotes and stellar isophotes (for example, in NGC 720) as indications of the presence of a significant dark halo (\citealt{Buote_canizares_disagniment_galaxies}; but see the following studies by \citealt{Buote_2002} and \citealt{Humphrey_NGC720}). Similar investigations had been performed also in the context of clusters of galaxies (for 5 Abell clusters; see \citealt{Buote_caniz_clusters}). On the other hand, it is not clear whether the assumption of hydrostatic equilibrium for the hot plasma is justified. In fact, attempts have been made recently at determining the conditions for a direct measurement of an appreciable rotation in the ICM (\citealt{ICM_rotation}, \citealt{Liu_Tozzi_2019}).\\
\indent If we now come to the focus of the present paper, three main facts set the dynamics of the galactic component of clusters of galaxies, as N-body systems, well apart from the dynamics of other gravitating systems (such as elliptical galaxies or globular clusters): the number of galaxies in clusters is rather small; some relevant time scales (in particular, dynamical time, classical relaxation time, and age) are often not well separated from one another; the galaxies are a minority component that contributes only little to the total gravitational field. Furthermore, especially for distant clusters, we have only recently achieved the capability of collecting accurate and reasonably complete kinematical data for their galaxies and of making an accurate measurement of the total gravitational field. 
Under these circumstances, it is natural that some questions related to the dynamical state of 
the galactic component, de facto already posed almost a century ago by \cite{Zwicky_ComaCluster} and \cite{Smith_VirgoCluster}, still remain unanswered.\\
\indent In particular, are there clusters in which galaxies are endowed with significant amounts of total orbital angular momentum? In other words, are there clusters where the galactic component exhibits significant mean rotation? Flattening is often caused by rotation, but we know that in general bright ellipticals (collisionless stellar systems) are not flattened by rotation. In turn, recently it has been found that rather round globular clusters (weakly collisional stellar systems), such as M15, 47 Tuc, and (the basically collisionless) $\omega$ Cen (e.g., see the \citealt{Rotating_globul_clusters}), when examined in detail show clear evidence of rotation. What causes the flattening of the galactic component of clusters, observed so frequently?\\
\indent After the pioneering work of the 1930s (\citealt{Smith_VirgoCluster} drew his conclusions on the absence of mean rotation in the Virgo cluster based on only thirty-two kinematical data points), early studies on the galactic component of nearby rich clusters tended to rule out the presence of internal mean motions (e.g., for the highly flattened cluster A2029 see \citealt{Dressler_norot}). Actually, in some cases indications of rotation were claimed. The possibility of slow rotation for the Coma cluster was noted by \cite{Tifft_coma_rotation} and for SC0316-44 by \cite{Materne_rotating_cluster}. A recent survey performed by \cite{Hwang_Lee} on the kinematics of the galactic component in 56 rich Abell clusters (starting from a survey on the galactic membership of 899 clusters) found 12 systems candidates to be in significant rotation; here the significance of rotation was defined by a threshold set arbitrarily by the authors. Curiously, for these 12 clusters the position angles of the minor axes of the isophotes of the visible component distributions do not appear to correlate with the position angles of the identified rotation axes.\\
\indent Other open questions (largely independent of the issue of rotation) refer to the relaxation state of the galactic component of clusters. Presumably, galaxies should be treated as a collisionless component (see for example \citealt{morph_classification} and \citealt{Sarazin_book}). Given their apparent quasi-equilibrium state, we do not actually know whether their phase space properties generally conform to the paradigm of incomplete violent relaxation (see \citealt{Violent_relax_LyndenBell}; \citealt{VanAlbada1982}) or for some reason they happen to conform to properties naturally expected in more collisional systems. Given the rather small number of galaxies that are involved, we lack rigorous decisive theoretical arguments in favor of one definite scenario and thus we may prefer to look for answers by means of a direct empirical approach. In particular, collisional systems are expected to evolve toward a dynamical state characterized by isotropic velocity dispersion, energy equipartition, and mass segregation. In contrast, collisionless systems should behave differently. Are there empirical signs that make the galactic component of clusters closer to a collisional stellar system or rather to the products of collisionless collapse?\\
\indent Of course, some differences in the galactic populations of clusters may have a different origin, related to processes of formation and mergers, and need not be associated with the dynamics of N-body systems and the role of galaxy-galaxy collisions in the context of classical relaxation. In any case, a clarification of the general dynamical state of the galactic component is desired. Further support for this last statement is obtained if we recall that several morphological classifications of galaxy clusters have been proposed, mainly based on the properties of the galactic component, as seen projected along the line-of-sight. Clusters can be arranged on a scale ranging from regular to irregular. For example, \cite{classification_zwicky} classified clusters as compact, medium compact, or open, based on the concentration of galaxies projected on the sky. \cite{classification_bautz} gave a classification system based on the degree to which the cluster is dominated by its brightest galaxies. \cite{morph_classification} reported no evidence of segregation in regular clusters at magnitudes fainter than 2 mag below the BCG. The irregular, spiral-rich clusters showed no evidence of mass segregation at any of the studied magnitude intervals.\\
\indent The objective of this paper is to study the dynamical state of the galactic component of two clusters, Abell S1063 and MACS J1206, at $z \approx 0.4$, for which accurate and radially extended photometric and spectroscopic 
observations are available. The structure of the two clusters has been studied previously in great detail. Their visible components are significantly flattened. In one case, the observed kinematics has been argued to reflect the presence of a recent merger. 
Otherwise, the two systems appear to be characterized by remarkable regularity. 
In particular, as illustrated by Fig. 2 in \citet{Bonamigo_alignment}, the
distributions of the DM, ICM and galactic components are very
well aligned with each other and associated with a well-defined common center.\\
\indent Here we focus on two specific points: the possible presence of mean rotation and selected empirical signatures of the type of relaxation that characterizes their quasi-equilibrium state, i.e. mass segregation, energy equipartition, and velocity dispersion anisotropy. 
We find evidence of mean rotation and energy equipartition, but we do find significant evidence of mass segregation. 
As to the velocity space distribution, we resort to two different continuum descriptions within idealized spherically symmetric models. First, we examine the available galaxy line-of-sight velocity distribution by means of the Jeans equations, and obtain indications on the relevant anisotropy dispersion profile. Because the qualitative behavior that is found is similar to that occurring in stellar systems formed by incomplete violent relaxation, we then proceed to apply a second method of investigation, by assuming that the galactic  component can be described by a physically justified distribution function, which in other contexts is known to incorporate the main features of the process of collisionless violent relaxation. Here the novel aspect of the method is that the total gravitational potential in the distribution function is provided empirically and imposed as an external potential on the basis of the modeling carried out separately for the two clusters (using gravitational lensing information). The two different approaches support each other and confirm that the phase space properties of the two clusters are qualitatively similar to those that characterize the products of incomplete violent relaxation.\\
\indent This paper is organized as follows. 
In Section 2, we present the photometric and spectroscopic data used in our investigation. 
In Section 3, we summarize some basic properties of the galactic component, in particular we describe the projected density and line-of-sight velocity dispersion profiles of the two clusters. 
In Section 4, we test the possible presence of mean rotation in three ways. 
In Section 5, we examine the possible presence of energy equipartition and (stellar) mass segregation. 
In Section 6, we apply the two dynamical models, of the Jeans equations and of a specific form of the galaxy distribution function, to fit the available data. 
Discussion and conclusions are presented in Section 7. \\

Throughout this paper, we adopt $H_0 = 70$ km s$^{-1}$ Mpc$^{-1}$,
$\Omega_{\rm m} = 0.3$,  $\Omega_\Lambda= 0.7$. At the clusters redshift, 1 arcmin corresponds
to 0.29 Mpc (AS1063, $z=0.35$) and 0.35 Mpc (M1206, $z=0.44$).

\section{Data sample}
\begin{figure*}
\centering
\includegraphics[width=\linewidth]{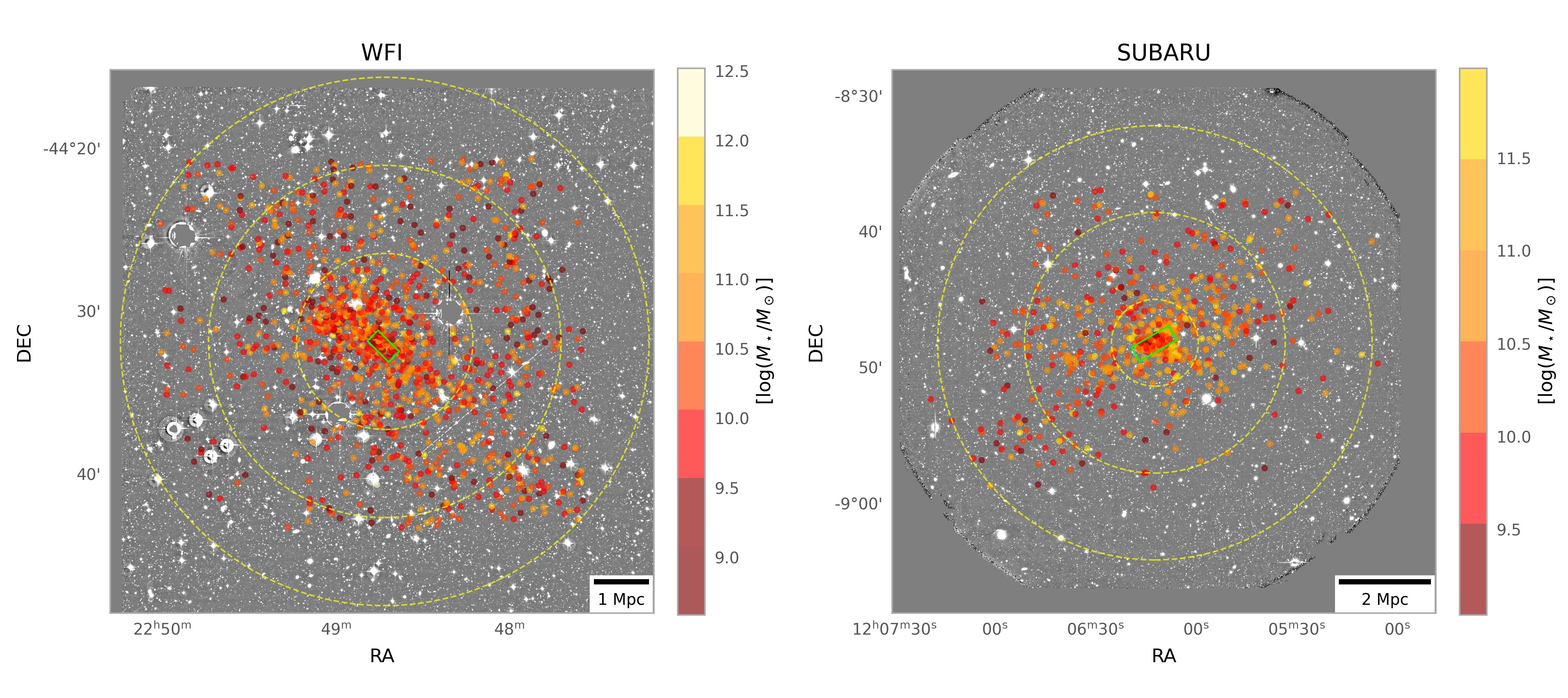}
    \caption{A view of the cluster members of Abell S1063 (left) and MACS J1206.2$-$0847 (right). The grey-scale background image on the left (right) is the WFI (SUBARU) observation in the $R_c$ band, spanning an area of $\approx 25\times25$ arcmin$^2$.
    The member galaxies are plotted as colored dots. The colors correspond to the stellar mass values of each galaxy as reported in the colorbar.
    The three concentric yellow circles represent the distances of 1.5, 3, 4.5 Mpc (1, 3 and 5 Mpc) from the cluster center, identified with the BCG.
    The green rectangle represents the $\approx 1\times2$ arcmin$^2$ ($\approx 1\times2.63$ arcmin$^2$) area covered by the MUSE spectroscopic follow-up on the inner regions of the cluster.}
    \label{fig:FITS_clusters}
\end{figure*}

The data used in this paper rely mostly on the outcome of the CLASH-VLT project, presented
in \citet{CLASH_VLT}.
The CLASH-VLT program builds upon the Cluster Lensing And Supernova survey with
Hubble (CLASH) project \citep{Postamn_CLASH}, a panchromatic, photometric
survey that collected observations from the near-ultraviolet to the near-infrared of 25 massive
clusters of galaxies at a median redshift of $z \approx 0.4$.
The CLASH-VLT program performed a spectroscopic campaign with the VIMOS instrument, a wide-field imager and multi-object spectrograph \citep{VIMOS}, mounted on the Very Large Telescope VLT to identify cluster members and lensed multiple images associated with the sub-sample of 13 clusters from CLASH visible from the Atacama sky.\\
\indent The CLASH-VLT campaign was significantly enhanced and complemented
in the cluster core by the observations collected with the integral field spectrograph
MUSE \citep{MUSE}.
This dataset offers the opportunity of trying to answer some of the still open questions discussed in this paper, since it contains as much information as currently available on the dynamics of the galactic component of clusters of galaxies. \\
\indent The two clusters analyzed in this work, Abell S1063 and MACS J1206.2$-$0847, were chosen because they are some of the best studied objects in the CLASH-VLT sample, and they both have additional observations with other spectrographs.
Both clusters have large-field photometry in the $R_c$ band, performed with the Wide Field Imager (WFI) at the MPG/ESO telescope \citep{WFI_imager} for Abell S1063 and with the Suprime-Cam \citep{Subaru_suprime_cam} on SUBARU telescope \citep{Umetsu_CLASH_M1206} for MACS J1206.2$-$0847. \citet{Bonamigo_alignment} showed that the DM, ICM, and galactic components of both clusters are almost perfectly concentric (within $\approx 10$ kpc).\\
\indent Abell S1063 is an intermediate symmetry I-II Bautz-Morgan type \citep{Abell_catalogue_1989}. MACS J1206.2$-$0847 is found to be an optically very rich system with a single dominant central galaxy and no obvious sub-clustering \citep{Temperature_M1206}, making it likely a Bautz-Morgan type I cluster. \\
\indent Fig. \ref{fig:FITS_clusters} shows a view of the member galaxies of Abell S1063 and MACS J1206.2$-$0847.

\subsection{Abell S1063 data}
Abell S1063 (also known as RXC J2248.7$-$4431, labeled as AS1063) was first observed
by \citet{Abell_catalogue_1989} and subsequently studied in detail as part of the CLASH and CLASH-VLT campaigns and of the Hubble Frontier Fields program by \citet{Frontier_Fields}. AS1063
is a massive cluster (\citealt{Sartoris_R2248} estimated a virial value of about $2 \times 10^{15}M_\odot$) at redshift $z = 0.35$. 
Although the cluster has a regular shape, \citet{Abell_merger} and \citet{Abell_merger21} found from X-ray flux analyses that it may have recently undergone an off-axis
merger. 
This hypothesis is challenged by the concentric distributions of the different mass components found by \citet{Bonamigo_alignment} (see also \citealt{Beauchesne_AS1063} for an independent X-ray and strong lensing analysis). 
The ICM in this cluster reaches temperature values of $k_B T \approx 12$ keV. AS1063 is an optimal gravitational lens that presents a set of 55 spectroscopically confirmed multiple images from 20 background sources, distributed over a wide range of redshifts ($0.73 < z_\text{source} < 6.11$, see \citealt{Caminha_A1063}).
The CLASH-VLT spectroscopic campaign yielded 3607 reliable redshifts, measured over a region of $\approx 25\times25$ arcmin$^2$, 1109 of which are classified as cluster members according to techniques that use the projected phase-space distribution of galaxies around the median redshift value $z = 0.3457 \pm 0.0001$ of the cluster (\citealt{Mercurio2021}). 
The catalog of cluster members and multiply lensed sources resulting from MUSE observations in the $1 \times 2$ arcmin$^2$ central region was presented in \citet{Caminha_A1063}, along with the first cluster strong lensing model. 
The full catalog was presented in \cite{Mercurio2021}. 
The MUSE data provided 175 new secure redshifts, 104 of which are classified as cluster galaxies. 
By considering 21 additional redshifts from the literature (for more information about the data sample, see section 2 in \citealt{Sartoris_R2248}), a total of 3850 spectroscopic redshifts were analyzed and 1234 cluster members were selected for the dynamical analysis presented in that work (see also \citealt{Mercurio2021}). 
Additional photometric data, collected through the WFI, were necessary to measure the values of stellar mass of the cluster member galaxies from composite stellar population models. 
\REV{In the following analysis, we will use the set of 1215 spectroscopically confirmed member galaxies with a stellar mass estimate.} 
\REV{Following \citet{Mercurio2021}, we inferred the stellar mass of the member galaxies from the information on the spectral energy distribution (SED) given by the available multiband photometry, using composite stellar population models based on \cite{Bruzual_Charlot} templates. 
The stellar mass values are estimated by assuming a \citet{Salpeter_IMF} stellar Initial Mass Function (IMF), a delayed exponential star formation history, solar metallicity and the presence of dust according to \cite{CalzettiASPC2000}.}
The radial completeness of our spectroscopic sample is discussed in \citet{Mercurio2021} and summarized in Table \ref{table:AS1063_completness}. The completeness of the sample is defined as the percentage of the cluster members with apparent magnitudes in the range $18.0 < m_{R_c} < 23.0$ which were spectroscopically observed during the survey. 
The completeness decreases with the projected distance from the cluster center.
\REV{In Appendix B, we test the robustness of our sample by running the same analysis on the subsample of cluster members with apparent magnitudes in the range $18.0 < m_{R_c} < 22.0$.}

\subsection{MACS J1206.2$-$0847 data}
MACSJ1206.2$-$0847 (M1206, hereafter) was observed with the HST during the CLASH survey, and then with the VLT/VIMOS as part of the CLASH-VLT program. 
The full redshift VIMOS catalog of MACS1206 consists of 3292 objects, with measured redshifts mostly acquired as part of our ESO Large Programme 186.A-0798 (P.I.: Piero Rosati). 
Additional archival VIMOS data have been homogeneously reduced from programs 169.A-0595 (P.I.: Hans Böhringer) and 082.A-0922 (P.I. Mike Lerchster). 
M1206 is a massive galaxy cluster (\citealt{Biviano_13} estimated a virial mass of $1.4\times10^{15}M_\odot$) at redshift $z = 0.4398 \pm 0.0002$ (\citealt{Girardi_M1206}). 
The cluster has a regular shape, presents a bright BCG and its ICM has a temperature of $k_B T \approx 11.6$ keV (see \citealt{Temperature_M1206}).
\REV{A combined analysis of strong and weak lensing, X-ray surface brightness and temperature, and the Sunyaev-Zel’dovich effect (see \citealt{Sereno_M1206} and \citealt{Chiu_M1206}) showed that the gas component of M1206 is remarkably regular and in hydrostatic equilibrium.}
M1206 is an excellent gravitational lens that presents 82 spectroscopic multiple images belonging to 27 families spanning the redshift range $1.0 < z_\text{source} < 6.1$ (see \citealt{Caminha_M1206}).
For this cluster, we have merged two datasets taken from \citet{Biviano_13} and \citet{Caminha_M1206}. \citet{Biviano_13} obtained a final catalog containing 2749 objects with reliable redshifts. Starting from this, the authors adopted a combination of two algorithms, the “Peak + Gap” method from \citet{Fadda_P_G} and the “Clean” method from \citet{Mamon_CLEAN}, to identify the spectroscopic members of the cluster. They obtained a sample of 592 member galaxies (see also \citealt{Annunziatella_M1206} for more information on this dataset).
Additional 141 galaxies observed with VLT/MUSE in the inner regions of the
cluster were spectroscopically confirmed as members of M1206 (see \citealt{Caminha_M1206}).
By merging these two catalogs we have found that 44 members were in common between the two, thus providing a final total number of member galaxies of 689.
Photometric data in five bands ($B$, $V$, $R_c$, $I_c$, $z'$) obtained with SUBARU were available for 658 of the 689 spectroscopically confirmed members. 
\REV{We used this multiband photometric data to extended the catalogue of 497 of these cluster members stellar masses obtained in \citet{Annunziatella_M1206}, adopting their fitting technique on the SED performed with the code MAGPHYS \citep{MAGPHYS}.}
This code is based on the stellar population synthesis models by \citet{Bruzual_Charlot}, with a \citet{Chabrier_IMF} stellar IMF\REV{, a continuum star formation rate with superimposed random bursts, metallicity values between 0.02 and 2 solar, and the dust model of \cite{Charlotte_Fall_2000}}.
We have converted the stellar mass values from a Chabrier to a Salpeter IMF \citep{Chabrier_to_Salpeter}, to properly compare the stellar mass distributions of the two clusters.
The completeness of the VIMOS sample is discussed in \citet{Biviano_13}, while the galaxy sample observed with the MUSE instrument is approximately 100\% complete in the magnitude range $18.0 < m_{R_c} < 23.0$.

\begin{table*}
    \centering
        \begin{tabular}{ llll }
        \hline \hline
          Radial interval & $0 < \frac{R}{(1 \text{ Mpc})}<  0.25$ & $0.25 < \frac{R}{(1 \text{ Mpc})}<  1$ & $1 < \frac{R}{(1 \text{ Mpc})}<   2.75$ \\
         \hline
         Completeness     & 100\%                    & $\approx 80$\%              & $\approx 75$\%              \\
         \hline \hline
        \end{tabular}
    \caption{Completeness of AS1063 spectroscopic sample. The radial completeness of the spectroscopic sample up to $R_c = 23.0$ mag, more details are given in \cite{Mercurio2021}. The typical error on the completeness percentage is $\approx5$ \%.}
    \label{table:AS1063_completness}
\end{table*}

\section{Basic properties of the galactic component}

\begin{figure}
  \centering
  \includegraphics[width=1.\linewidth, keepaspectratio]{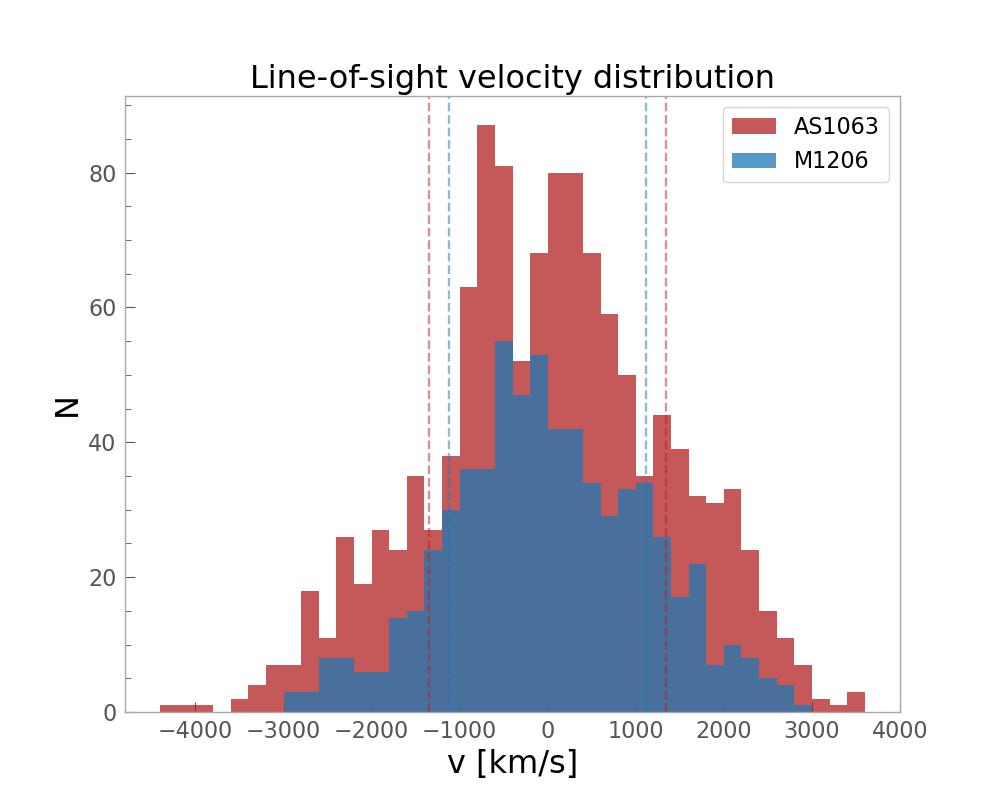}
  \caption{Line-of-sight velocity distribution for the member galaxies of the two clusters. The vertical dashed lines represent the corresponding velocity dispersion $\sigma_\text{los}$. The binning procedure consisted in dividing the range of observed l.o.s. velocities in \REV{bins of 200 km/s}.}
  \label{fig:vel_distribution}
\end{figure}

\begin{figure}
    \centering
    \includegraphics[width=1.\linewidth, keepaspectratio]{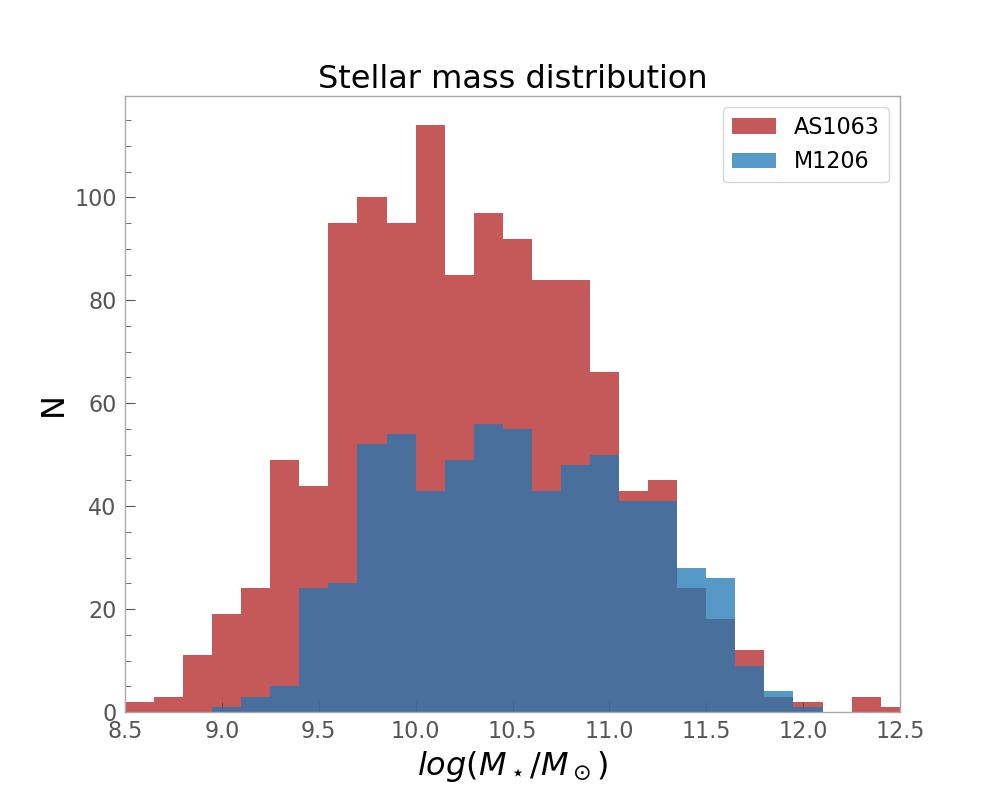}
    \caption{Stellar mass distribution for the two clusters\REV{, divided in bins of 0.15 dex}. Because the individual galaxies contain also dark matter haloes and gas, the total mass distribution would be shifted to the right of the diagram, possibly with a different shape.  }
    \label{fig:stel_mass_distribution} 
\end{figure}

\begin{figure}
  \centering
  \includegraphics[width=\linewidth, keepaspectratio]{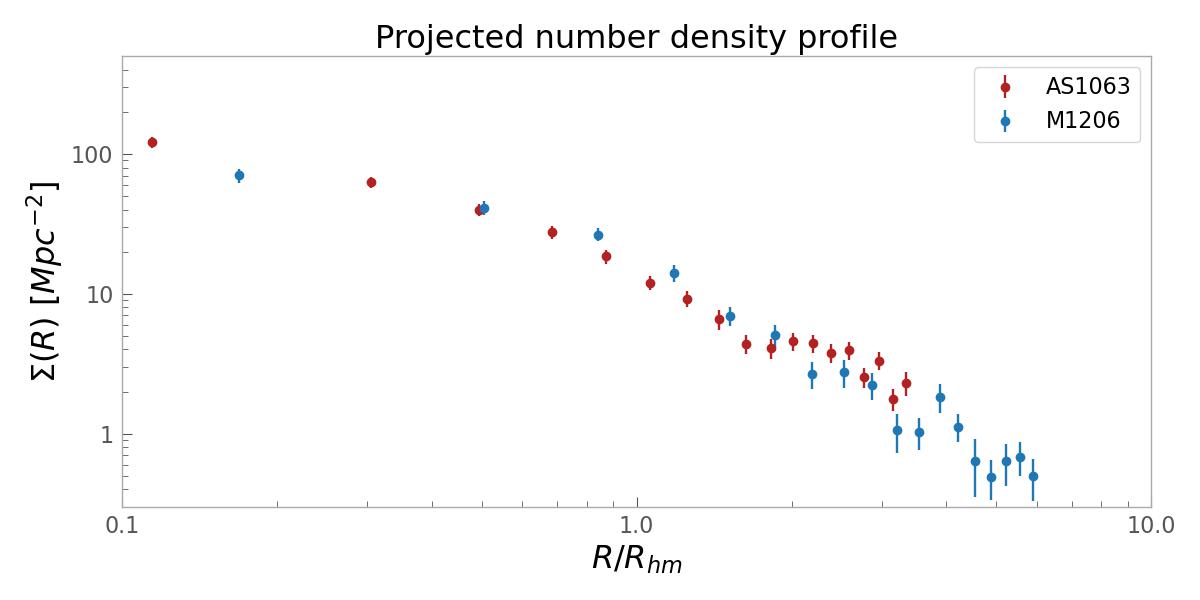}
  \caption{Projected number density profile of the galactic component as a function of the radial distance from the cluster center. For each cluster, the circularized radius $R$ is measured in terms of the half-mass radius defined on the basis of its stellar mass distribution. The bins are obtained by dividing in 20 sub-intervals the range of observed $R/R_{hm}$ of each cluster. The error bars are $1\sigma$. As briefly described in the text, the profiles are affected by \REV{the} sample \REV{in}completeness in a non-trivial way.}
  \label{fig:radial_surface_number_density}
\end{figure}

\begin{figure}
  \centering
  \includegraphics[width=\linewidth, keepaspectratio]{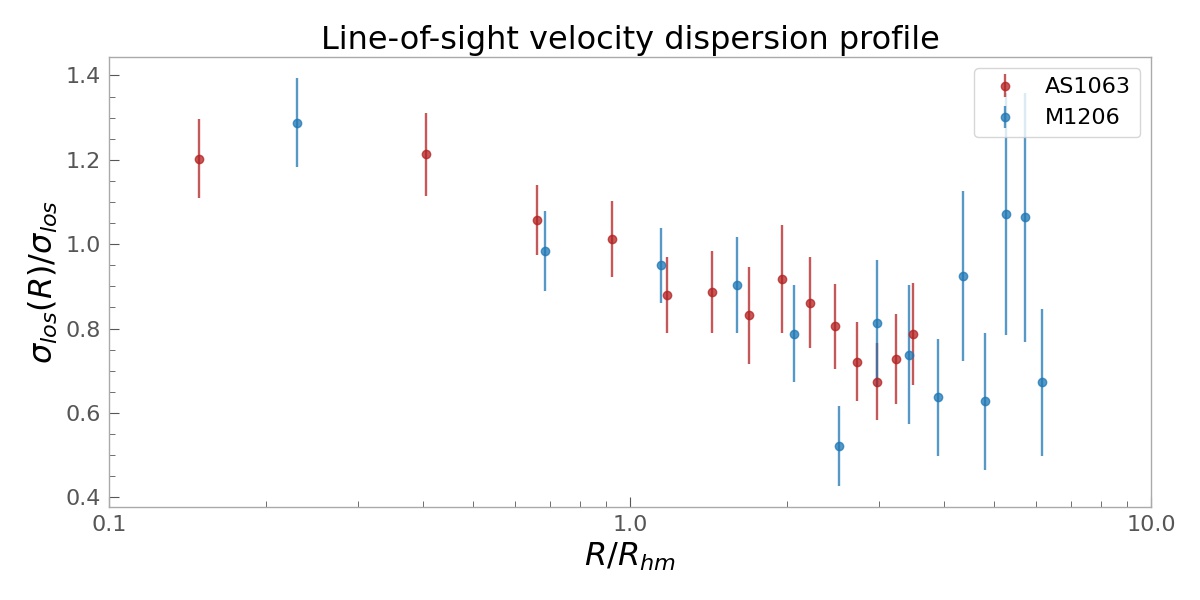}
  \caption{Line-of-sight velocity dispersion profile of the galactic component as a function of the radial distance from the cluster center, normalized to the global velocity dispersion $\sigma_\text{los}$ of each cluster. As in Fig. \ref{fig:radial_surface_number_density}, for each cluster the circularized radius $R$ is measured in terms of the cluster stellar half-mass radius, and the bins are defined according to the same procedure. The vertical bars represent the errors evaluated  as $\sigma_\text{los}(R)/\sqrt{N(R)}$, where $\sigma_\text{los}(R)$ is the velocity dispersion relative to the radial bin located at $R$ and $N(R)$ is the number of galaxies inside such bin.}
  \label{fig:radial_velocity}
\end{figure}

From the data presented in the previous section, we derive some
basic properties of the galactic component of the two clusters, in particular
the line-of-sight velocity and stellar mass distributions. In addition, after a 
standard circularization, we derive the related surface number density and velocity dispersion profiles.\\
\indent The velocity distributions for the two clusters are plotted in Fig.\ref{fig:vel_distribution}. 
The velocity distribution of AS1063 exhibits two peaks that are $\approx 1000$ km/s apart. The separation of the two peaks could indicate the presence of overall rotation or the existence of two different structures undergoing a merging phase; it might also just be an artifact related to adopted binning. 
To test whether the two peaks in the velocity distribution of AS1063 are related to a bimodal distribution, we performed the so-called dip test, developed by \cite{dip_test_unimod}, which measures unimodality in a sample by the maximum difference over all sample points between the empirical distribution function and the unimodal
distribution function that minimizes that maximum difference.
The dip test gives a p-value of 0.976, which suggests that the distribution
is compatible with a single Gaussian distribution. Other tests would thus be 
required to study the possible presence of rotation (see Sect. 4) or merging. Furthermore, the dip test encourages us to consider the standard deviation of a Gaussian as a good choice for the global velocity dispersion even for the case 
of AS1063, apparently associated with a two-peaked velocity distribution profile. The velocity dispersions for AS1063 and M1206 are $\sigma^{AS1063}_\text{los} \approx (1350\;\pm\;26)$ km/s and $\sigma^{M1206}_\text{los} \approx (1120\;\pm\;29)$ km/s, respectively. 
For AS1063, \cite{Mercurio2021} obtained a velocity dispersion value of $1380^{+26}_{-32}$ km/s, which is compatible with our estimate.
For M1206, \cite{Girardi_M1206} obtained a velocity dispersion value of $1035^{+27}_{-45}$ km/s, which is lower than our result. 
The difference can be explained by the inclusion in our catalog of 97 additional sources detected with VLT/MUSE (\citealt{Caminha_M1206}) in the inner regions of the cluster, as mentioned in Section 2.\\
\indent The stellar mass distributions are plotted in Fig. \ref{fig:stel_mass_distribution}. The distributions underestimate the number of low-mass galaxies because of the observational limits in apparent magnitudes characterizing the sample.\\
\indent A preliminary step required in order to calculate the number density and line-of-sight velocity dispersion profile is the definition of the cluster center. 
For the two clusters, the distributions of galaxies, hot ICM, and dark matter are almost concentric (see \citealt{Bonamigo_alignment}, who found that the corresponding centers are within $\approx 10$ kpc from one another and basically coincide with the center of the cluster BCG). 
This is why we choose as the center of the two clusters the position in the sky of their BCGs, which is well constrained by the photometric data available. 
The fact that the three main mass components are concentric provides strong observational evidence that the clusters have reached a well-defined quasi-equilibrium state.\\
\indent For simplicity, we assume that the galactic projected number density distribution in the two clusters has contour levels described by confocal ellipses with constant eccentricity, defined in terms of the major and minor semi-axes $a$ and $b$ as $e = \sqrt{1 - b^2/a^2}$. 
The eccentricities of the two clusters are listed in Table \ref{table:summary}. By converting these values into ellipticities ($\varepsilon = 1 - b/a$), we note that the flattening that is involved is similar to that of E3-E4 elliptical galaxies. We recall that the eccentricity of a discrete distribution of points can be found from the second moments of such distribution (for the derivation and an application to galaxy clusters, see \citealt{ellipticities}). Then, in constructing the number density and velocity dispersion profiles we refer to the circularized radius $R = \sqrt{ab}$, in which the projected distance of a galaxy from the cluster center is defined in terms of the major and minor semi-axes of the isodensity contour on which the galaxy is located.\\
\indent The plots provided in Figs. \ref{fig:radial_surface_number_density} - \ref{fig:radial_velocity} are profiles expressed as a function of the dimensionless radius $R/R_{hm}$. Here the half-mass radius $R_{hm}$ is defined as the \REV{(circularized)} radius that encloses half of the stellar mass projected on the sky. Remarkably, in this representation the surface number density and velocity dispersion profiles of the two clusters basically overlap out to $\approx 3$ $R_{hm}$.\\
\indent Figure \ref{fig:radial_surface_number_density} illustrates the surface number density profiles. The observed surface number density can be approximated by a monotonically declining power law, at least out to $\approx 3 R_{hm}$, with exponent $\approx 1.5$ . 
This is consistent with the findings of \cite{Biviano_13} for M1206 (see Fig. 7 of that paper) and of \cite{Sartoris_R2248} for AS1063 (see Fig. 3 of that paper).
Figure \ref{fig:radial_velocity} shows the velocity dispersion profiles. The value of the velocity dispersion in each radial bin is normalized to the global velocity dispersion of the cluster considered. This is also consistent with the velocity dispersion profile found by \cite{Biviano_13} for M1206 (see Fig. 3 of that paper).\\
\indent In the cosmological context, a commonly used scale for the radius of a cluster is the so-called virial radius, generally denoted as $r_{200}$. Note that the ratio of $r_{200}$ to $R_{hm}$ for the two clusters is in the range $1.1 - 1.6$ (see Table \ref{table:summary}). Therefore, we can state that the similarities between the radial profiles of the two clusters represented in Figs. \ref{fig:radial_surface_number_density} - \ref{fig:radial_velocity} hold out to about $2 - 3$ virial radii.\\
\indent 
\REV{Figure \ref{fig:radial_velocity} shows that both clusters display a regular, monotonically decreasing velocity dispersion profile within 3 half-mass (i.e., $\approx 2$ virial) radii. This regularity seems in contradiction with a possible recent merger in AS1063 (\citealt{Abell_merger}, \citealt{Abell_merger21}, \citealt{Mercurio2021}, \citealt{Beauchesne_AS1063}). The kinematic data on the cluster members of M1206 extend beyond this point and it appears that the decreasing trend is interrupted, which may be interpreted as an indication that, in those outer regions, the cluster has not yet reached a quasi-equilibrium state.}\\
\indent In Table \ref{table:summary} we report an estimate of the relaxation time $T_D$ for the two clusters considered in this paper, based on the definition (see \citealt{Chandrasekhar_relax_time})

\begin{equation}\label{eq:relax_time}
T_D = \frac{\sigma^3}{8 \pi G^2 m^2 n \ln{\Lambda}},
\end{equation}
\noindent
where we have taken the standard estimate $\ln{\Lambda} \approx 10$ and for $\sigma$, $n$, and $m$ we have inserted the values obtained from the previous data by applying the following definitions: $\sigma=\sigma_\text{los}$ is the global velocity dispersion and $m$ is the typical mass of a galaxy (taken to be the mean value of the stellar mass distribution in Fig. \ref{fig:stel_mass_distribution}, ignoring the DM halo and gas contribution to the total mass budget of a single galaxy, which are expected to be a factor $\approx 2$-$3$ greater than the stellar mass alone).
The average number $n$ density used to obtain the relaxation time was derived from a power law fit on the measured surface number density (the average is performed within the radial interval \REV{$[0.1, 1]$ half-mass radii}). By comparing these values to those of the crossing time $2 R_{hm}/\sigma_\text{los} \approx 2$ Gyr and to the Hubble time, we thus confirm the common picture that the galactic component in clusters should be considered as collisionless (see for example \citealt{Sarazin_book}).


\begin{table*}
    \centering
        \begin{tabular}{ cccccccccc }
         \hline \hline
               & \# members & $z$  & $e$   & $\sigma_\text{los}$ &  $M_{\star, tot}$   & $M_{200}$  & $r_{200}$ & $R_{hm}$ & $T_D$  \\
               & (1) & (2)  & (3)   & (4) & (5) & (6) & (7) & (8) & (9)  \\
         \hline
         AS1063 & 1215       & 0.35 & $0.77$ & $1350$ km/s    & $7.32 \times 10^{13} M_\odot$ & $2.17 \times 10^{15} M_\odot$ & $2.70$ Mpc & $1.98$ Mpc   & $ 2\times 10^3$ Gyr    \\
         \hline
         M1206 & 658         & 0.44 & $0.75$ & $1120$ km/s    & $5.38 \times 10^{13} M_\odot$ & $1.40 \times 10^{15} M_\odot$ & $1.97$ Mpc & $1.21$ Mpc   & $ 7\times 10^2$ Gyr    \\
         \hline \hline
        \end{tabular}
    \caption{Structural properties of the two clusters. (1) Number of galaxies for which position in the sky, velocity along the line-of-sight, and stellar mass are available, selected to belong to the cluster. (2) Average cluster redshift $z$, based on the values of the redshift of the selected member galaxies. (3) Adopted eccentricity, as defined in the text (see Sect. $3$). (4) Global velocity dispersion, based on the velocity distribution shown in Fig. \ref{fig:vel_distribution}. (5) Total stellar mass $M_{\star, tot}$ associated with the distribution illustrated in Fig. 3. (6) Virial mass $M_{200}$ and (7) virial radius $r_{200}$, taken from \cite{Sartoris_R2248} for AS1063 and from \cite{Biviano_13} for M1206. (8) Half-mass radius $R_{hm}$, defined as the circularized radius that encloses half of the stellar mass projected on the sky. (9) Relaxation time $T_D$, as defined in Eq. (\ref{eq:relax_time}).}
    \label{table:summary}
\end{table*}

\section{Search for the presence of mean rotation}

\begin{figure}
  \centering
  \includegraphics[width=\linewidth, keepaspectratio]{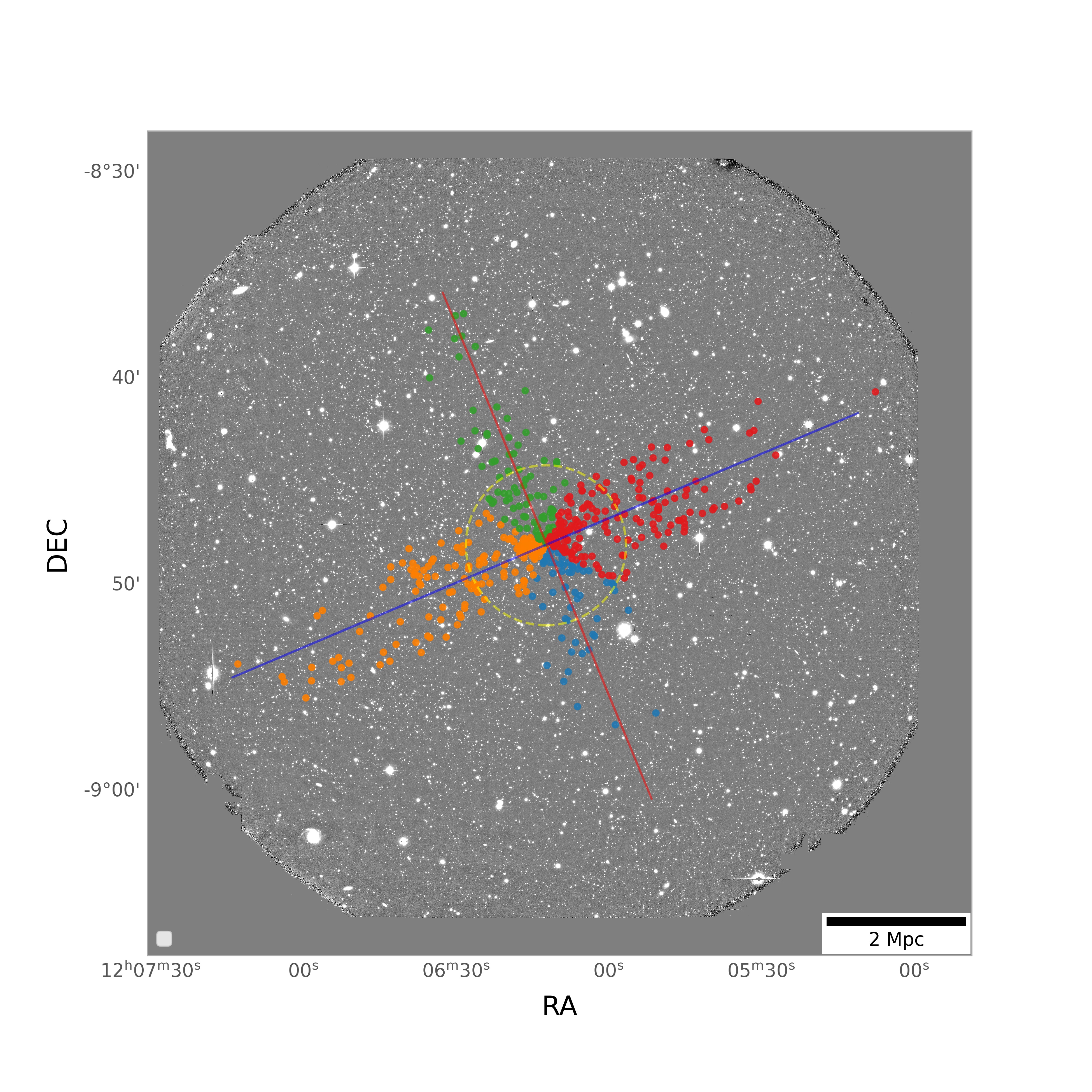}
  \caption{\REV{The sample of m}ember galaxies divided in sectors along the major (blue line) and minor (red line) axes for M1206\REV{, following the steps introduced in Sect. 4}. The dashed yellow circle represents the half-mass radius $R_{hm}$. The sectors are identified as sector A in blue, sector B in orange, sector C in green, and sector D in red.
  \REV{If rotation is responsible for the observed flattening, gradients in the line-of-sight galaxy velocity distribution would be expected along the major axes and not along the minor axes.}}
  \label{fig:MR_fin_M1206_rotating_selection_axis}
\end{figure}

\begin{figure}
  \centering
  \includegraphics[width=\linewidth, keepaspectratio]{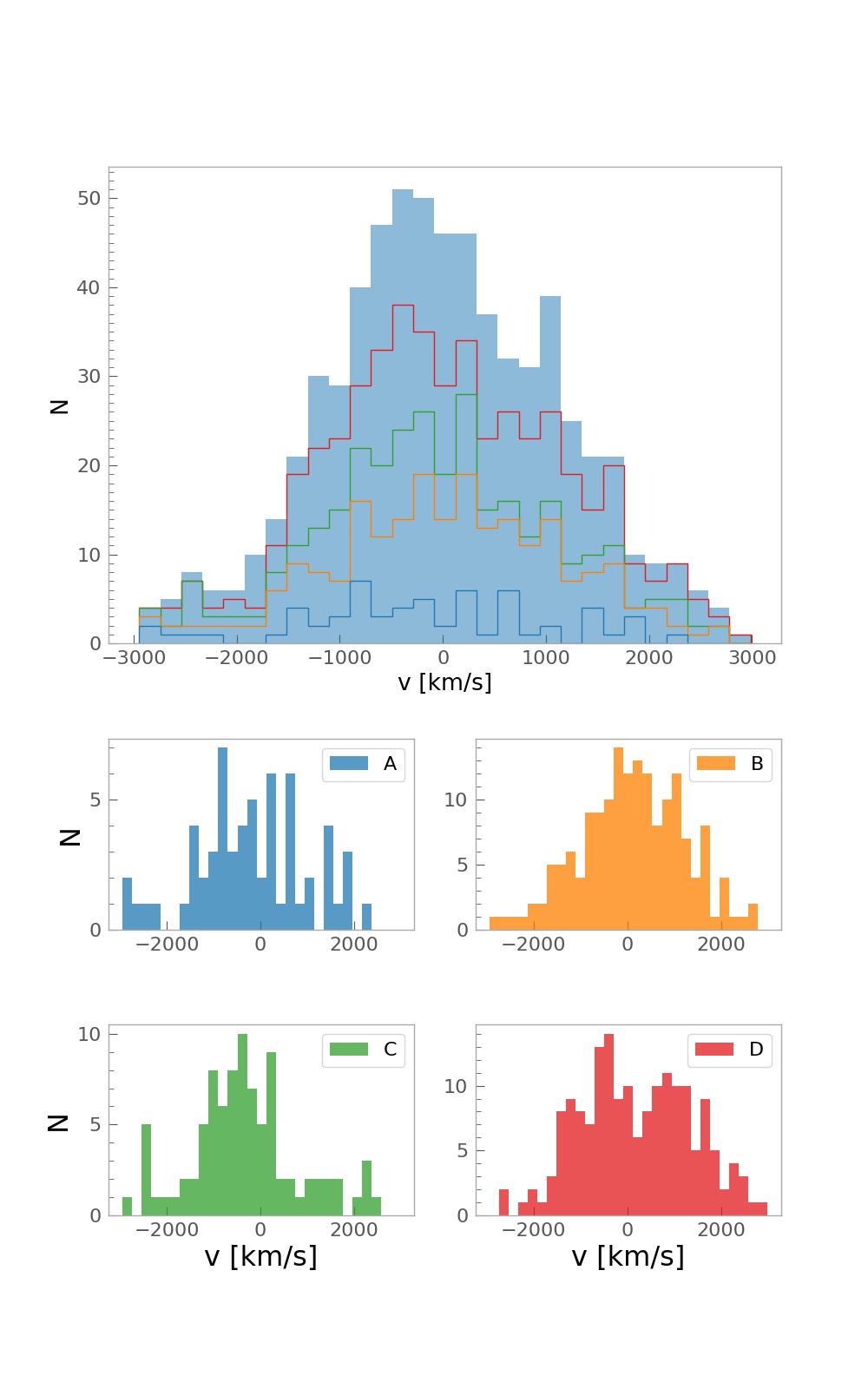}
  \caption{Distribution of the line-of-sight velocities of member galaxies of M1206 contained in the sectors shown in Fig. \ref{fig:MR_fin_M1206_rotating_selection_axis}. Stacked distribution of velocities in the four sectors compared to the total velocity distribution of Fig. \ref{fig:vel_distribution} (upper histogram). Velocity distributions of the galaxies in each sector (lower histograms). The sectors are identified as sector A in blue, sector B in orange, sector C in green, and sector D in red.}
  \label{fig:MR_fin_M1206_sector_vel_distribution} 
\end{figure}

\begin{figure}
    \centering
    \includegraphics[width=\linewidth, keepaspectratio]{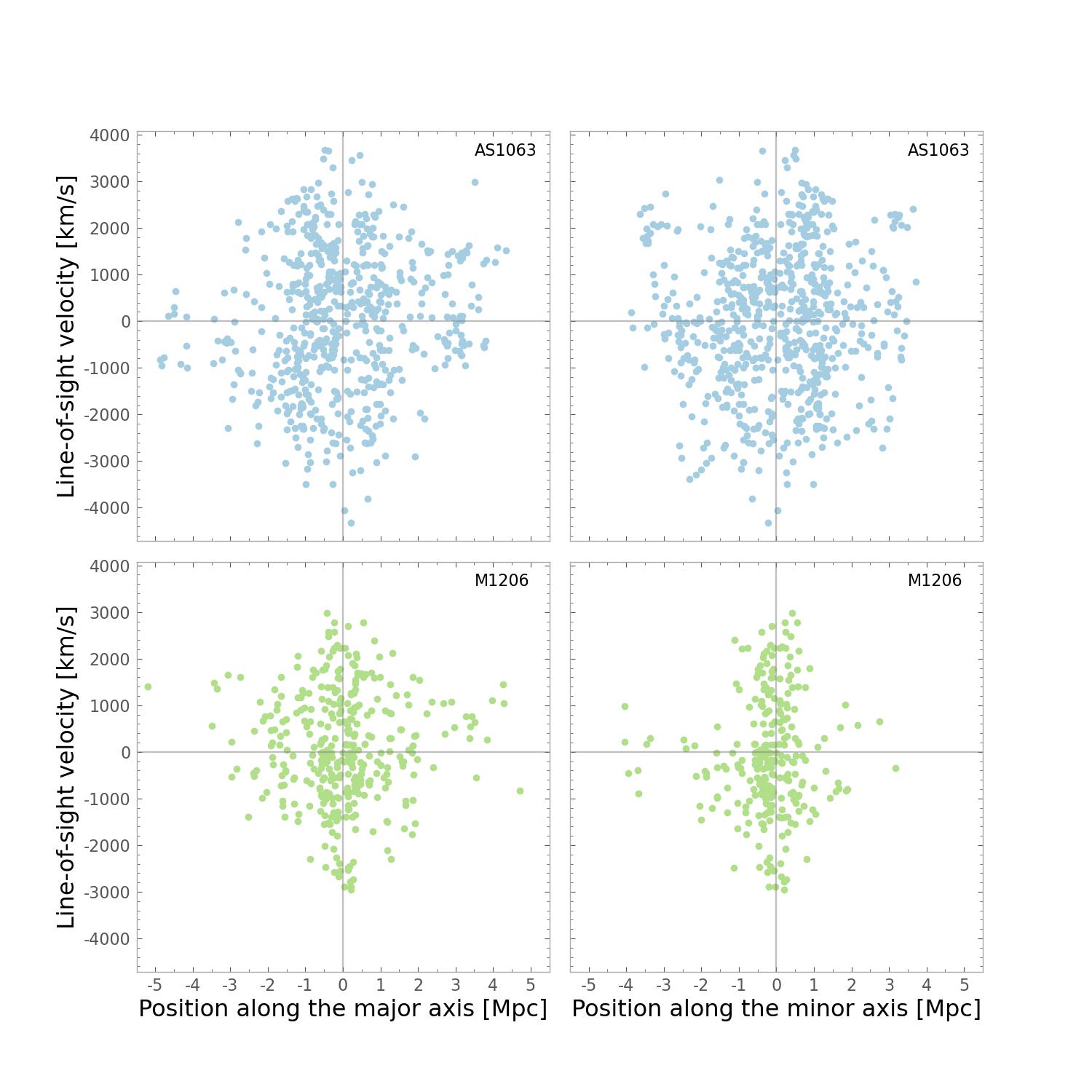}
    \caption{Position-velocity diagrams of the member galaxies within a distance $D<(1/2) R_{hm}$ from the \REV{major (left) and minor (right) axes} of AS1063 (cyan) and M1206 (green). These plots do not show any clear rotation signature (which would offset the points in the diagrams towards two diagonally opposite quadrants).}
    \label{fig:R_v_plot}
\end{figure}

In the previous section we noted that the spatial distribution of galaxies in the two clusters is flattened, similarly to E3-E4 galaxies. The principal axes of the galaxy distributions are aligned with those of the central BCG; in addition, the ICM and dark matter, as from X-ray and gravitational lensing analyses, are also characterized by flat spatial distributions, aligned with the galactic component (\citealt{Bonamigo_alignment}). 
If rotation is responsible for the observed flattening, gradients in the line-of-sight galaxy velocity distribution would be expected along the major axes, whereas negligible gradients should be found along the minor axes. 
In addition, as a reference value, we may refer to the $(V/\sigma, \varepsilon)$ diagram in Fig. 1 of the article by \cite{Davies_dyn_ellipticals} which suggests that for E3-E4 oblate isotropic rotators (following the very simple picture of the homogeneous self-gravitating classical ellipsoids as in \citealt{Chandrasekhar_Ellispoidal_FOE}) the expected ratio of peak rotation velocity $\abs{v_\text{rot}}$ to velocity dispersion should be $\approx 0.7$ \REV{(and $\approx 0.4$ for prolate rotators)}.
The quantity $V/\sigma$ is the ratio of the observed maximum of the rotation profile to the line-of-sight velocity dispersion.\\
\indent As a first simple test of the possible presence of rotation related to the observed flattening in the galactic component of AS1063 and M1206, we divided each cluster into four sectors in relation to its major and minor axes and then plotted the line-of-sight velocity distribution for the galaxies inside each sector. The sectors and the galaxy sub-samples for the two clusters were defined in the following way:
member galaxies with a projected distance from the cluster center $R < R_{hm}$ are divided in four circular sectors and grouped according to their closest principal axis, whereas galaxies with $R > R_{hm}$ are added to a sector if they are within a distance $D < (\sqrt{2}/2) R_{hm}$ from the closest principal axis. These sectors are highlighted in Fig. \ref{fig:MR_fin_M1206_rotating_selection_axis} for M1206.
For each cluster the line-of-sight velocities are the internal velocities, obtained after subtraction of the mean velocity of the full sample of galaxies as in Fig. \ref{fig:vel_distribution}. The results for M1206 are illustrated in Fig. \ref{fig:MR_fin_M1206_sector_vel_distribution} and some quantitative results for the two clusters are recorded in Table \ref{tab:vels}.\\
\indent We produced position-velocity diagrams for the cluster members found within a distance $D<(1/2) R_{hm}$ from the major \REV{and minor} axis, as shown in Fig. \ref{fig:R_v_plot}. 
Both tests summarized in Figs. \ref{fig:MR_fin_M1206_sector_vel_distribution} and \ref{fig:R_v_plot} suggest that there is little or no evidence for internal rotation\REV{, oblate or prolate (of the kind discussed in \citealt{Prolate_rot_Ebrova_Lokas}),} in the galactic component of the two clusters, when considering the whole samples.\\

\subsection{Rotation in the central region}

\begin{figure*}
  \centering
  \includegraphics[width=\linewidth, keepaspectratio]{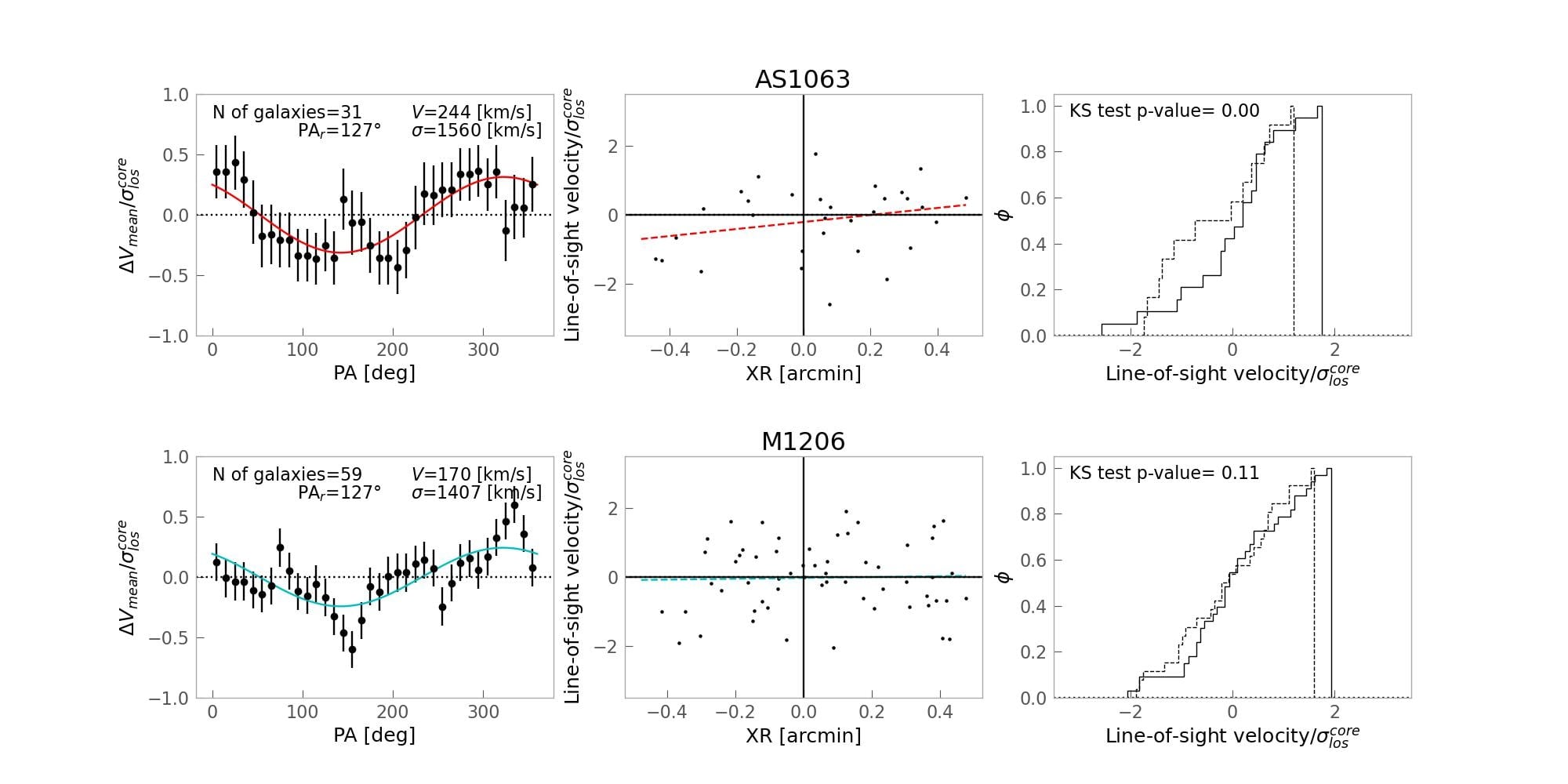}
  \caption{
  Diagnostic diagrams of the rotation signature detected in the core of the two clusters within $0.5'$ from the center (corresponding to the areas covered by the MUSE pointings). 
  The left panels show the difference between the mean velocities $\Delta V_\text{mean}$ on each side of a line passing through the center with the given angle (PA), divided by the velocity dispersion $\sigma_\text{los}^\text{core}$ in the core of each sample. 
  The continuous line is the sine function that best fits the observed patterns. 
  In these panels labels identify the number of galaxies within $0.5'$, the value of PA \REV{that maximizes $|\Delta V_\text{mean}|$ (PA$_{r}$)}, the rotation velocity $V$ (half of the amplitude) from the best-fit sine function, and the value of $\sigma_\text{los}^\text{core}$.
  The central panels show the position-velocity diagrams constructed by considering the line-of-sight velocities as a function of the projected distances from the rotation axis (XR) in arcminutes. 
  The dashed lines are the least square linear fits to the data. 
  The right panels show the cumulative line-of-sight velocity distributions for the galaxies with $XR<0$ (solid line) and for those with $XR>0$ (dotted line). 
  The labels provide the Kolmogorov-Smirnov p-value that the two samples are extracted from the same parent distribution. The p-value should be close to zero if the clusters are characterized by the presence of mean rotation.
  }
  \label{fig:central_rotation}
\end{figure*}

We should emphasize that the $(V/\sigma, \varepsilon)$ relation involves only global quantities (\citealt{Binney_Rotation_anis_1978}, \citealt{Binney_Rotation_anis_revisited_2005}).
The diagram shown in \cite{Davies_dyn_ellipticals} should be significantly modified to include inclination effects and the change in pressure anisotropy with radius (\citealt{Cappellari_Sauron_X}).
In addition, some self-gravitating stellar systems exhibit clear evidence of solid-body rotation only in their innermost regions (e.g., see \citealt{Rotating_globul_clusters}, \citealt{Leanza_2022_GC_rot}).\\
\indent To check whether the clusters have an isotropic (or mildly anisotropic) core characterized by significant mean rotation, we derived the rotation profile for the galaxies within \REV{a circular region of} $0.5'$ \REV{radius} from the center of each cluster 
(i.e., within the MUSE pointings, by maximizing the number of galaxies considered and thus avoiding a potentially sharp variation of the sample completeness that occurs at larger radii).\\
\indent We first identify the angle (PA) of the projected rotation axis in the plane of the sky (defined as the angle between the rotation axis and the right ascension axis). 
To identify the value of PA, we follow a standard procedure in globular cluster studies (e.g., see \citealt{Cote_1995}, \citealt{Rotating_globul_clusters}, \citealt{Leanza_2022_GC_rot}):
we compute the mean line-of-sight velocities in the two sub-samples obtained by dividing the cluster members by a line passing through the cluster center with a given PA. 
The PA is varied in steps of 10 degrees and the difference between the mean velocities $\Delta V_\text{mean}$ is plotted against PA.
The resulting pattern is fitted with a sine function. 
The PA at which the maximum difference in mean velocities is reached corresponds to the (projected) rotation axis and the amplitude of the sine function is twice the rotation amplitude $V$.
If the clusters are characterized by the presence of mean rotation, the position-velocity diagrams would show an asymmetry, and the galaxies on each side of the rotation axis are expected to be sampled from distinct cumulative velocity distributions.
The results of this analysis are plotted in Fig. \ref{fig:central_rotation}.\\
\indent Figure \ref{fig:central_rotation} indicates that in the core of both clusters there is significant evidence of rotation ($V/\sigma \approx 0.15$).
The minor axis of the galaxy distribution in the core is offset by 30 degrees from the best fit rotation axis.
The eccentricity (ellipticity) of AS1063 and M1206 within $0.5'$ from their centers is $e=0.60$ ($\varepsilon = 0.2$) and $e=0.77$ ($\varepsilon = 0.37$) respectively.
The ratio of peak rotation velocity to velocity dispersion required to sustain an edge-on oblate isotropic rotator with the ellipticity of AS1063 (M1206) would be ${V/\sigma}\approx 0.4$ ($\approx 0.6$).
However, by relaxing the condition on isotropy, the ratio of ${V/\sigma}$ required to maintain an oblate shape is lowered (\citealt{Cappellari_Sauron_X}). 
For example, a mild anisotropy (this concept and the related definition will be properly introduced in Sect. 6.2) of $\beta \approx 0.15$ would be compatible with the requirement of an oblate rotator with the shape of the core of AS1063.\\
\indent \REV{A prolate rotator (of the kind studied in \citealt{Binney_Rotation_anis_1978}) would require a lower amount of pressure anisotropy to sustain the observed eccentricity of the cores, although it would introduce serious modelling complications as the rotation axis is perpendicular to the symmetry axis and thus the component of the specific angular momentum along the rotation axis is not an integral of the motion.}\\
\indent\REV{As a further confirmation of a rotating core picture, we note that Fig. 22 in \cite{Sereno_M1206} shows a drop in the ratio of thermal and non-thermal pressure (possibly related to the presence of bulk motions, see \citealt{2010Meneghetti_bulkmotions}) in the inner $\approx 100$ kpc/$h$ of M1206, which corresponds exactly with the region within 5 arcmin where we detect the presence of mean rotation.}\\
\indent In Appendix C we apply the analysis discussed in this subsection to the whole sample, and to the galaxies within 5 arcmin, confirming the absence of rotation at large radii. 
As a side note, a test for rigid rotation similar to that considered by \cite{Hwang_Lee} fails to give a clear signal of rotation in the center for these two clusters (because of the low signal-to-noise ratio of the mean rotational motion), while it correctly indicates the absence of rotation at large radii.


\begin{table*}
    \centering
        \begin{tabular}{ |c|cccc|cccc| }
         \hline
         & \multicolumn{4}{c|}{AS1063} & \multicolumn{4}{c|}{M1206} \\
         \hline
                                                    & sector A & sector B & sector C & sector D             &sector A &  sector B      &  sector C  & sector D \\
         \hline
        $\langle{v}\rangle_\text{sec}$    [km/s]    &$-$3.6 & $-$26.9 & 91.0   & $-$199.1                   & $-$194.4 & 111.6 & $-$333.8 & 191.2\\
        $\sigma_\text{los, sec}$   [km/s]           &1357.2 & 1483.9  & 1420.0 & 1327.2                     &1189.6 & 1104.2 & 1146.1 & 1171.6\\
         \hline
        \end{tabular}
    \caption{Summary of the mean velocities $\langle{v}\rangle_\text{sec}$ and velocity dispersion $\sigma_\text{los,sec}$ inside each sector. The sectors are defined in Fig. \ref{fig:MR_fin_M1206_rotating_selection_axis} and the distribution of the line-of-sight velocities of member galaxies in each sector are represented in Fig. \ref{fig:MR_fin_M1206_sector_vel_distribution}. Sectors B and D are those relative to the major axes.}\label{tab:vels}
\end{table*}

\section{Empirical evidence of the absence of collisional relaxation}
\begin{figure}
    \centering
    \includegraphics[width=\linewidth, keepaspectratio]{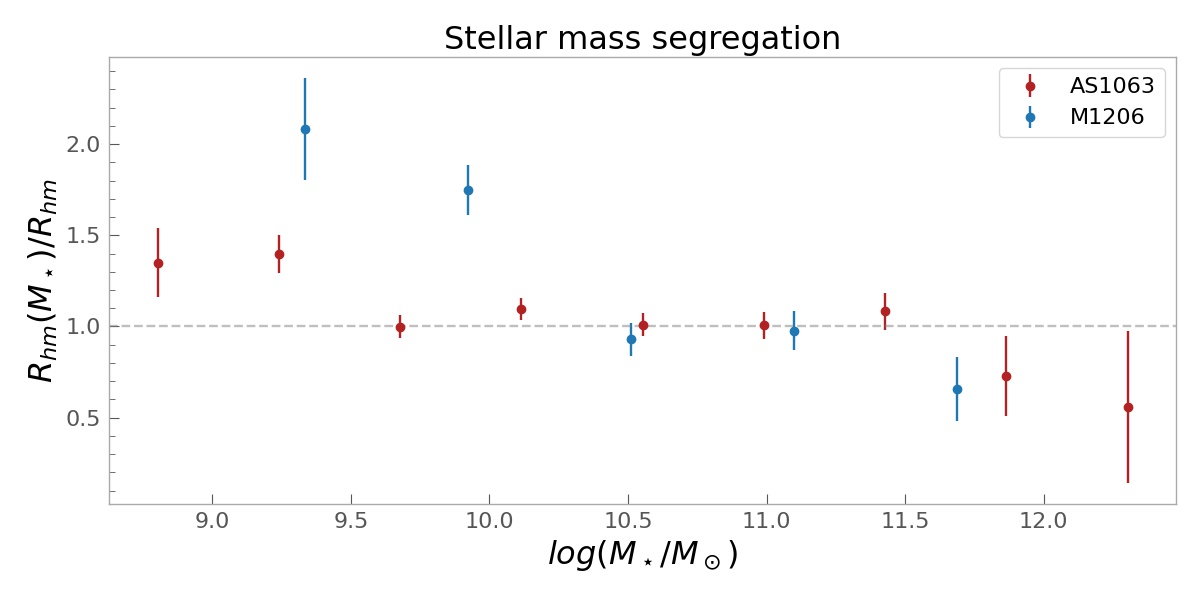}
    \caption{Evidence of mass segregation. The galaxies are divided in bins according to their stellar mass. Variation of $R_{hm}(M_\star)$, that is the half-mass radius of the cluster members inside the $n$-th bin normalized to the global half-mass radius, with galactic mass. The errors are evaluated as $s_N/\sqrt{N}$, where $s_N$ is the standard deviation of the projected radius distribution inside the $n$-th bin, and $N$ is the number of elements inside the $n$-th bin.}
    \label{fig:mass_segregation}
\end{figure}

\begin{figure}
    \centering
    \includegraphics[width=\linewidth, keepaspectratio]{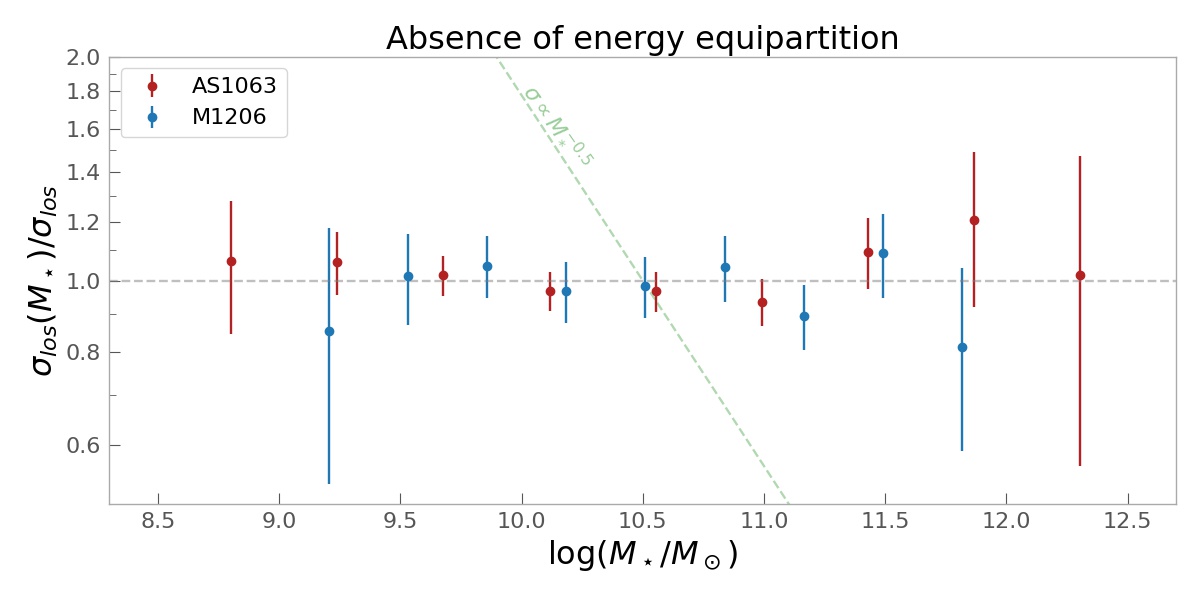}
    \caption{Absence of energy equipartition. The galaxies are divided in bins according to their stellar mass. Variation of $\sigma_\text{los}(M_\star)$, that is the velocity dispersion of the cluster members inside the $n$-th bin normalized to the global velocity dispersion, with galactic mass. The errors are evaluated as $\sigma_\text{los}(M_\star)/\sqrt{N}$, where $N$ is the number of elements inside the $n$-th bin. In green, the trend expected in the presence of complete energy equipartition ($\sigma_\text{los} \propto M_*^{-0.5}$)}
    \label{fig:energy_equipartition}
\end{figure}

In relaxed thermodynamical (i.e., collisional) systems, independently of the initial conditions, two properties are expected to be established: energy equipartition, that is the velocity dispersion of particles of mass $m$ should scale as $m^{-1/2}$, and mass segregation, that is lighter particles should be characterized by a more diffuse distribution than that of heavier particles \REV{(e.g., see \citealt{Smith_VirgoCluster}, \citealt{1987_Spitzer_GC})}.\\
\indent To look for hints of mass segregation, we divided the galactic component in mass bins based on the galaxy stellar mass $M_\star$. We then studied the ratio of the half-mass radius measured on the sub-distribution of galaxies associated with each bin to the global half-mass radius $R_{hm}$ as a function of $M_\star$. 
In Fig. \ref{fig:mass_segregation} we observe that indeed a decreasing trend for $R_{hm}(M_\star)$ is present in both clusters, which is more pronounced in M1206.
Therefore, both clusters appear to be mass-segregated, relative to the stellar mass.
Of course, in terms of total mass, mass-segregation is likely to be associated with a profile different from that of Fig. \ref{fig:mass_segregation}. Furthermore, the characteristics of the profiles illustrated in Fig. \ref{fig:mass_segregation} need not result from the slow action of collisionality, but rather from different evolutionary processes, such as galaxy merging.
\REV{At first glance, our results might seem at odds with those by \cite{Biviano_13}, who found that the concentration of the luminosity density is around 10\% higher than the concentration of the luminosity density, reporting little evidence for mass segregation.
This cumulative approach ignores the stellar mass distribution shown in Fig. \ref{fig:stel_mass_distribution}.
The less common massive galaxies can be segregated and still give a relatively small contribution to the overall concentration.}
\\
\indent In any case, mass segregation poses a contradiction to the simple use of one-component models in the interpretation of the observations. This point has already been addressed for M1206 by \cite{Biviano_13}, who considered separately the populations of passive and star-forming galaxies (which roughly correspond to elliptical and spirals respectively). In a different physical context, to take into account the possible presence of mass segregation two-component models have also been considered to describe globular clusters (for example, see \citealt{fnu_globular_cluster} and \citealt{Torniamenti}).\\
\indent In Fig. \ref{fig:energy_equipartition}, we compare the velocity dispersion for galaxies with different stellar masses, adopting the same binning procedure as the one used in Fig. \ref{fig:mass_segregation}. This plot suggests that AS1063 and M1206 do not exhibit evidence of energy equipartition. Further analyses focusing only on the central regions ($R<(1/5)$ $R_{hm}$) confirm the absence of energy equipartition.\\
\indent The absence of equipartition, in spite of the presence of some mass segregation, supports the picture that the galactic component of the two clusters should be considered a collisionless component and that the origin of mass segregation should reflect some evolutionary process in the formation of the two clusters unrelated to galaxy-galaxy scattering.


\section{Dynamical models}

\subsection{Number density from the observations}

In the next subsection, we will apply the Jeans equations to make dynamical predictions for the velocity dispersion profiles, testing different hypotheses on the velocity dispersion anisotropy present in the two clusters. To do so, we will assume that the number density of galaxies is known (from the available number counts) and that the overall potential well is known (from published gravitational lensing investigations). 
\REV{In the context of stellar dynamics, spherically symmetric models often proved successful in describing astrophysical objects with a wide variety of apparent morphologies (e.g. see \citealt{SagliaBertinStiavelli1992} Sect. 3.2, and in particular Fig. 1, for a test on the effectiveness of spherical models on strongly flattened - E4/E6 - elliptical galaxies).
On the other hand, the task of trying to take into account the dynamical effects of flattening or triaxiality would be non-trivial and the required enormous complexity of such modeling would not be sufficiently justified by the available data.}
For simplicity, we will develop our models under the assumption that each cluster is spherically symmetric.\\
\indent In this context, one key step is the derivation of the volume (three-dimensional) number density $n(r)$ starting from the surface (projected) number density $\Sigma(R)$ of the galactic component of the two clusters\footnote{Given the incomplete information available on the \REV{galaxy mass distribution} for the two clusters, we will refer to the number density $n$ instead of the mass density $\rho$.}:
\begin{equation}\label{eq:surface_projection_cap_6}
    \Sigma(R) = \int_{-\infty}^{\infty} n(r)\dd z = \int_{R}^{\infty} n(r) \frac{2r \dd r}{\sqrt{r^2-R^2}},
\end{equation}
where $r$ is the three-dimensional radius and $R$ is the radius projected on the sky surface.
Deriving the volume number density from the projected density $\Sigma(R)$ corresponds to the classical problem known as Abel inversion. Here we prefer to work on the direct problem, starting from three well-known spherically symmetric analytical power law volume density profiles, namely the \cite{Jaffe_prof} profile, the NFW profile (see \citealt*{NFW_profile}) and the spherical limit of the dual Pseudo Isothermal Ellipsoid (dPIE, see \citealt{dPIE_profile}). These profiles depend on two scales (two scales and one dimensionless parameter for the dPIE), that is a density scale and a length scale.

\begin{figure*}
  \centering
  \includegraphics[width=\linewidth, keepaspectratio]{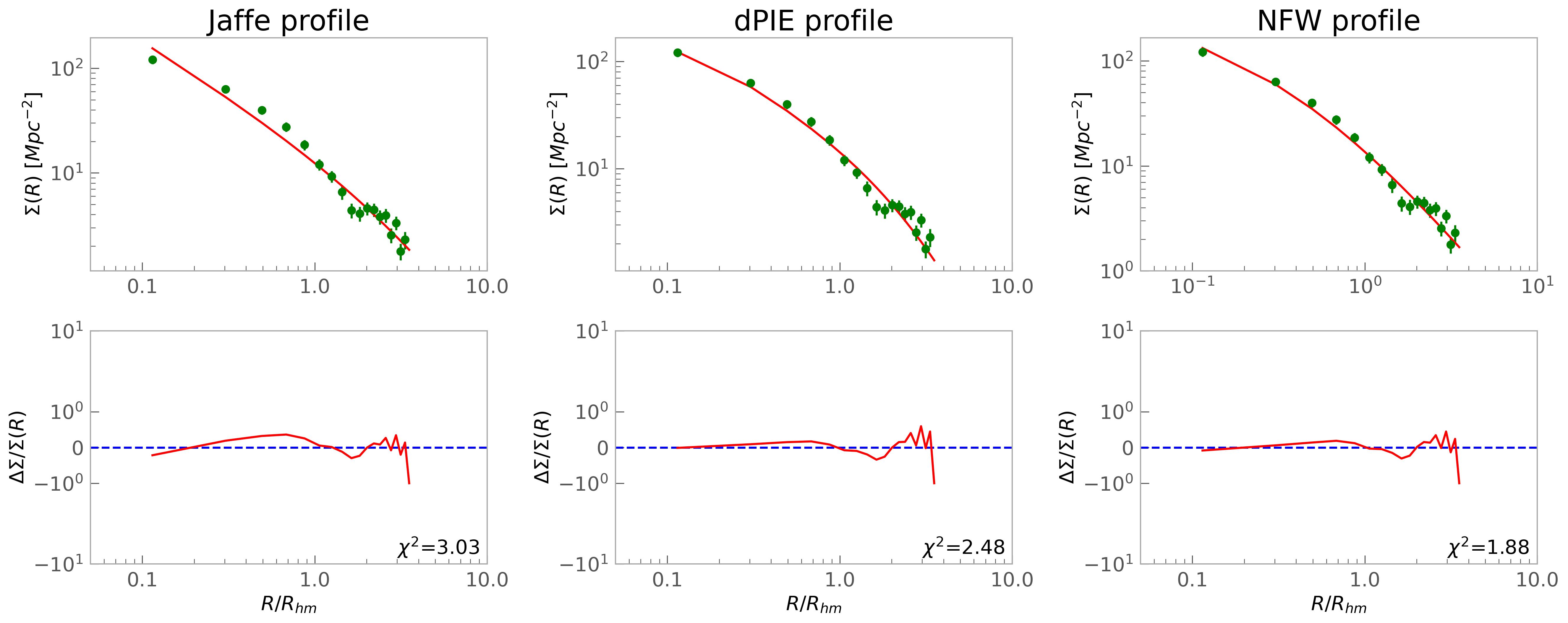}
  \caption{The observed surface number density of AS1063 (in green) fitted with the projected Jaffe, dPIE and NFW profiles (in red). 
  Top: the profiles resulting from the best-fit process.
  Bottom: the fractional deviations between the analytical model and the photometric data. \REV{The labels provide the reduced $\chi^2$ values.}}
  \label{fig:fit_surfdensity}
\end{figure*}

The Jaffe number density profile is
\begin{equation}\label{eq:Jaffe_profile}
    n_{J}(r) =  \frac{n_0}{(r/r_s)^2(1+r/r_s)^2}
\end{equation}
 The NFW number density profile is
\begin{equation}\label{eq:NFW_profile}
    n_{NFW}(r) = \frac{n_0}{(r/r_s)(1+r/r_s)^2}
\end{equation}
The dPIE number density profile is
\begin{equation}\label{eq:dPIE_profile}
    n_{dPIE}(r) = \frac{ n_0}{[1+(r/r_{core})^2][1+(r/r_{cut})^2]}
\end{equation}
with $r_{cut} > r_{core}$.\\
\indent The Jaffe profile was developed in the context of the dynamics of elliptical galaxies; close to the origin, it approaches the profile of a singular isothermal sphere, and at large radii is characterized by a decline that guarantees a finite total number of particles.\\
\indent The NFW profile is often adopted in the cosmological context, because numerical simulations on the growth of dark matter halos appear to be well modeled by this profile.\\
\indent The dPIE profile is mostly used in gravitational lensing studies to model the total mass distribution of a lens because of its convenient analytical properties; it is approximately constant below $r_{core}$ and scales approximately as $n(r)\sim r^{-2}$ in the radial range $r_{core}<r<r_{cut}$.\\
\indent These profiles admit a simple analytical expression for the projected density profile, that is for the integral of Eq. (\ref{eq:surface_projection_cap_6}) (e.g., see \citealt{Jaffe_proj} for the Jaffe profile and \citealt{NFW_dPIE_Projection} for the dPIE and NFW profiles).\\
\indent We fitted the observed projected number density profiles shown in Fig. \ref{fig:radial_surface_number_density} by these simple models. For the cluster AS1063 the results are illustrated in Fig. \ref{fig:fit_surfdensity} and show that the three models all fit reasonably well the available photometric data.

\subsection{Jeans equations}

Under different assumptions on the local anisotropy profile $\beta(r)$ defined as a function of the diagonal elements of the velocity dispersion tensor $\sigma_{ii}$ as
\begin{equation}\label{eq:beta}
\beta(r) = 1-\frac{1}{2}\frac{\sigma_{\theta \theta}^2 + \sigma_{\phi \phi}^2}{\sigma_{rr}^2}
\end{equation}
\noindent
we will test whether the Jeans equations (for assigned galaxy number density profile and total gravitational field in the two clusters AS1063 and M1206) predict line-of-sight velocity profiles for the galactic component compatible with the observations. For simplicity, the analysis will be carried out under the assumption of spherical symmetry, absence of mean rotation, and proportionality between number and mass density for the galactic component. The relevant (radial) equation can be written as

\begin{equation}\label{eq:Jeans_spheric_cap_6}
\frac{1}{n(r)}\dv{(n(r) \: \sigma_{rr}^2(r))}{r} + \frac{2\:\beta(r)\:\sigma_{rr}^2(r)}{r} = - \frac{G M(r)}{r^2}
\end{equation}
The boundary condition adopted to solve this differential equation is defined at large radii, where the velocity dispersion $\sigma_{rr}(r)$ is assumed to vanish. Here the function $M(r)$ is the total mass enclosed in a sphere of radius $r$, which is dominated by the contributions of dark matter and ICM; for this we refer to the final models based on strong and weak gravitational lensing described in \cite{Umetsu_strong_weak_lens_CLASH} (see Table \ref{tab:umetsu_nfw_totalmass}).
The total mass profile for a NFW model expressed as a function of the parameters listed in Table \ref{tab:umetsu_nfw_totalmass} is
\begin{equation}\label{eq:total_mass_profile}
M(r) = M_{200} \frac{\ln\left(1+r/r_s\right)-r/(r+r_s)}{ \ln\left(1+c_{200}\right)-c_{200}/(1+c_{200})}
\end{equation}
The number density profiles are taken from the analysis of Subsect. 6.1; we will refer in  the following only to the representation in terms of a Jaffe profile. \REV{The choice of a Jaffe profile is motivated by its regular properties (i.e. finite total mass). 
In Appendix D we show the results of the Jeans analysis using a NFW number density profile, which turn out to be very similar.}\\
\indent By projecting the solution for the radial velocity dispersion we obtain
\begin{equation}\label{eq:sigma_los_jeans}
\sigma_\text{los}^2 (R) = \frac{2}{\Sigma(R)}\int_R^\infty \left(1-\beta(r)\frac{R^2}{r^2}\right)\frac{n(r)\:\sigma_{rr}^2(r) \:r \:\dd r}{\sqrt{r^2-R^2}}
\end{equation}
for the predicted velocity dispersion along the line-of-sight (the integration we implemented follows Appendix 2 of \citealt{Mamon_Lokas_Jeans_Kernel}.).

\begin{table}
  \centering
    \begin{tabular}{ cccc }
     \hline \hline
     cluster          &  $M_{200}$ [$10^{14} M_\odot$]    &  $c_{200}$    &  $r_s$ [Mpc]   \\
     \hline
     AS1063            &  $18.78\pm6.72$               &$3.6\pm1.4$  & $0.66\pm0.32$\\
     \hline
     M1206            &  $18.17\pm4.23$               &$3.7\pm1.1$  & $0.60\pm0.21$\\
     \hline \hline
    \end{tabular}
  \caption{Cluster parameters derived for the total mass distribution described by a spherical NFW model, reconstructed from combined strong-lensing, weak-lensing shear and magnification measurements. $M_{200}$ and $c_{200}$ are respectively the virial mass and concentration parameter (e.g., the ratio between the virial radius $r_{200}$ and the NFW scale radius $r_s$). The fit to the relevant data was performed out to 3 Mpc for the two clusters. From Table 3 in \cite{Umetsu_strong_weak_lens_CLASH} (there the cluster AS1063 is listed as RX J2248.7$-$4431).}
  \label{tab:umetsu_nfw_totalmass}
\end{table}

\begin{figure}
  \centering
  \includegraphics[width=\linewidth, keepaspectratio]{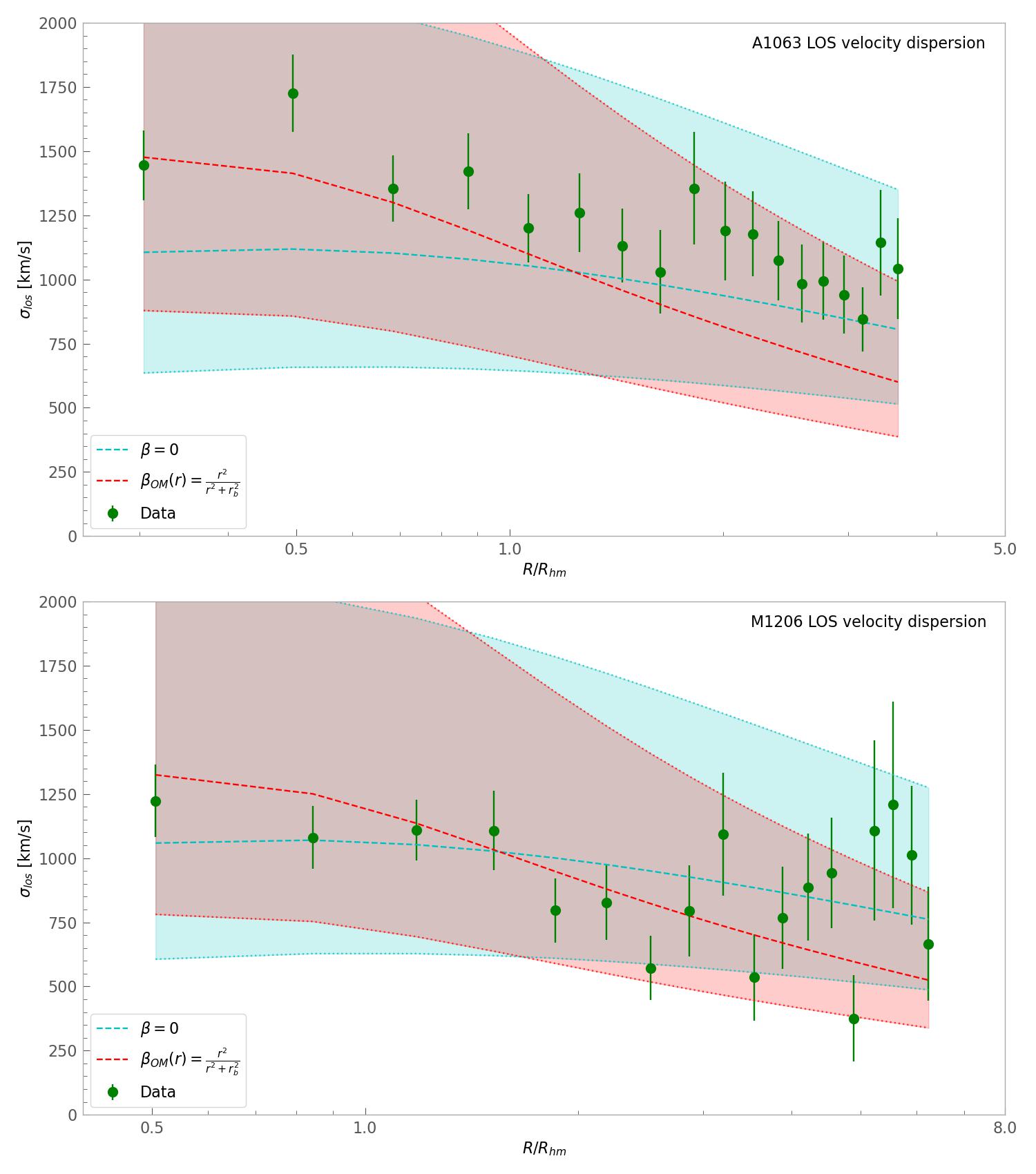}
  \caption{The line-of-sight velocity dispersion profile of the galactic component obtained from the Jeans equation for AS1063 and M1206. The plots show in green the data points derived from the spectroscopic data (see Fig. \ref{fig:radial_velocity}), superimposed on the velocity profiles derived from the Jeans equation, for two choices of $\beta(r)$: in cyan the constant isotropic $\beta=0$ profile and in red the Osipkov-Merritt anisotropy profile. The confidence regions are found by allowing the total mass $M(r)$ to vary within the uncertainties on the values of its parameters reported in Table \ref{tab:umetsu_nfw_totalmass}. The confidence levels are underestimated for $R>$3 Mpc because in this region the total mass $M(r)$ is extrapolated and not directly constrained by the observations.}

  \label{fig:Jeans_vel_disp}
\end{figure}

For the local anisotropy profile we decided to take the following three-parameter function
\begin{equation}\label{eq:beta_aniso_prof_general}
\beta(r)= \beta_\infty\left(\frac{r^\gamma}{r^\gamma+ r^\gamma_\beta}\right)
\end{equation}
The function $\beta(r)$ thus exhibits an isotropic (i.e., $\beta  =  0$) center and may be either radially ($\beta>0$) or tangentially ($\beta<0$) biased at large radii ($r > r_\beta$). The set of parameters considered in our
analysis are listed in Table \ref{tab:beta_param}.\\
\indent The results of the Jeans equation approach are illustrated in Fig. \ref{fig:Jeans_vel_disp}. For an easier reading, it shows the velocity dispersion profiles associated only with two choices of  $\beta(r)$, the isotropic ($\beta_\infty = 0$, $r_\beta = 0$) and the so-called Osipkov-Merritt (OM) profile ($\beta_\infty = 1$, $\gamma = 2$). While the confidence regions of both solutions can reproduce the kinematical data reasonably well, it is clear that the decreasing trend of the observed velocity dispersion profile is better represented in the solution associated with the OM anisotropy profile (in particular within 3 Mpc, which is the range within which the gravitational field was obtained by \citealt{Umetsu_strong_weak_lens_CLASH}).\\
\indent This is consistent with other results present in the literature, especially in the work by \cite{Biviano_13} on M1206 (in particular, see Fig. 15 of this article) and by \cite{Sartoris_R2248} on AS1063 (see Table 1 of this article).\\
\indent This velocity dispersion anisotropy profiles suggests the presence of a mild anisotropy in the core ($\beta(0.5')\approx0.05$), slightly less than the amount required to justify the shape of the core.

\begin{table}
  \centering
    \begin{tabular}{ cccccc }
     \hline \hline
                      &  iso-   &  const.       &  const.               & radially  & Osipkov     \\
                      &  tropic &  radial       &            tangential &    biased & -Merritt     \\
     \hline
     $\beta_\infty$   &  0           &1/2                & $-$1/2               & 1/2             & 1               \\
     \hline
     $r_\beta$  [Mpc] &  0           &0                  & 0                  & 1               & 1               \\
     \hline
     $\gamma$              &  0           &0                  & 0                  &  1              & 2               \\
     \hline \hline
    \end{tabular}
  \caption{ The set of parameters describing the velocity dispersion anisotropy profile of Eq. (\ref{eq:beta_aniso_prof_general}). The parameter $\beta_\infty$ represents the value of the anisotropy in the outer regions, $r_\beta$ is a scale length, and $\gamma$ sets the steepness of the profile.}
  \label{tab:beta_param}
\end{table}

\subsection{A dynamical model based on a physically justified distribution function}

In this last subsection, we explore the possibility that the galactic component might be described by a distribution function of the form that was originally conceived to incorporate the picture of formation by incomplete violent relaxation (in particular, see \citealt{F_nu_stiavelli}, \citealt{Bertin_Stivelli}, \citealt{Trenti_bertin}). In the context of the dynamics of individual elliptical galaxies, the function has been successfully used to model the observations not only in the one-component case for galaxies in which dark matter appears to be lacking (as is the case of NGC 3379\REV{; see \citealt{SagliaBertinStiavelli1992}, \citealt{BertinStiavelli_Report}, and \citealt{Romanowsky_noDM})}, but also in self-consistent two-component models for galaxies in which evidence of the presence of dark matter is significant (as is the case of NGC 4472; see \citealt*{SagliaBertinStiavelli1992}). Here the exploration is attempted in a completely different situation, that is, the case in which the component described by the distribution function (i.e., the galaxies of the two clusters) does not contribute significantly to the total gravitational field in which it is embedded. In other words, here the mean gravitational potential that will appear in the distribution function is assumed to be an external potential, due entirely to the ICM and dark matter components.\\

\begin{figure}
    \centering
  \includegraphics[width=\linewidth, keepaspectratio]{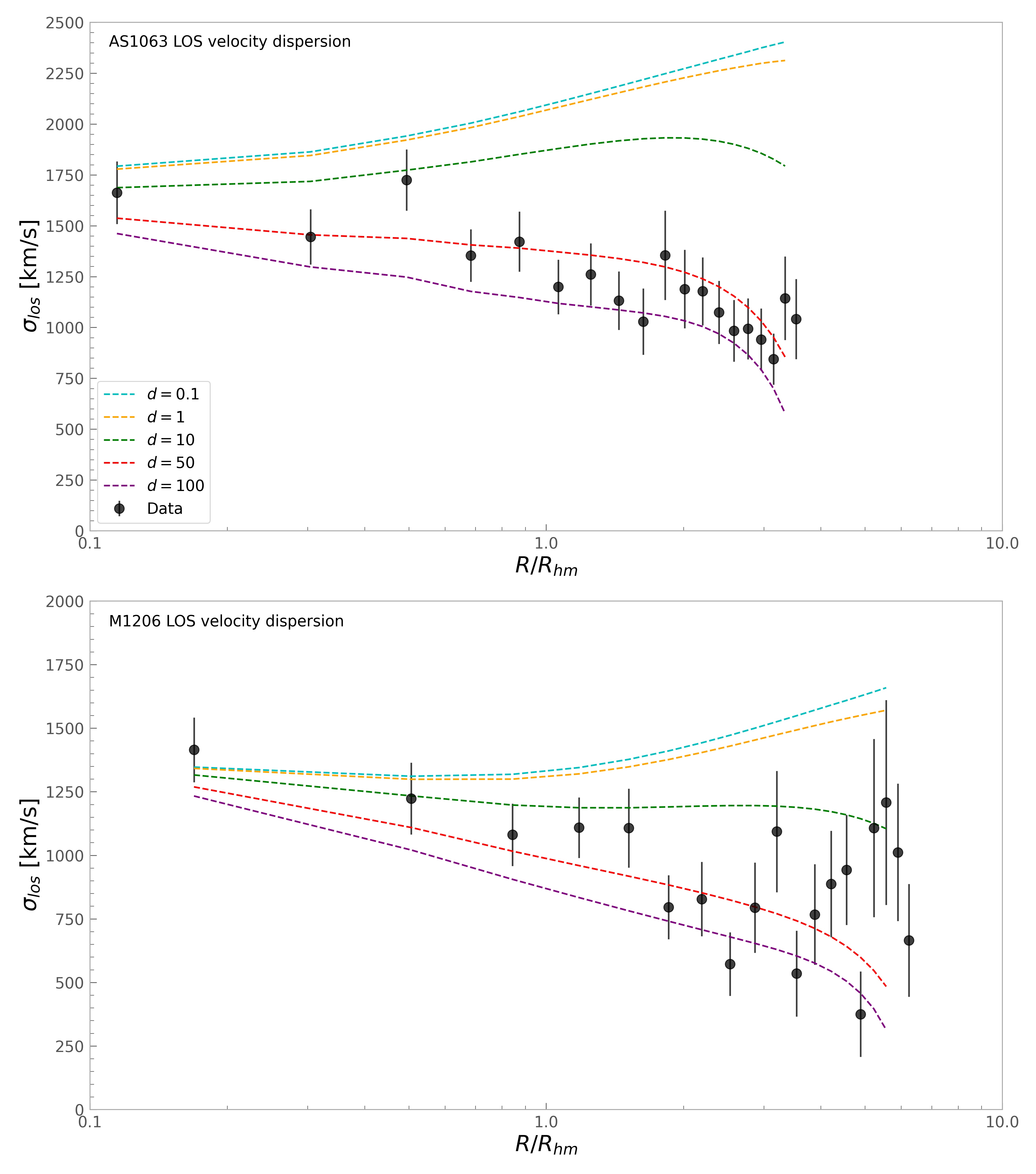}
  \caption{The line-of-sight velocity dispersion profile of the galactic component derived from the $f^{(\nu)}$ distribution function with fitted parameters for AS1063 and M1206 (see Table \ref{table:summary_cap_6_5}). 
  The plot shows in black the data points derived from spectroscopic data (see Fig. \ref{fig:radial_velocity}), superimposed on the velocity profiles derived from the moments of the distribution function, for 5 choices of the value of the parameter $d$ (see Fig. \ref{fig:DF_beta} for the associated anisotropy profile).
  Both clusters are well described by the velocity dispersion curves associated to a value of $d\approx 10$-$50$.}
  \label{fig:DF_vel_dispersion}
\end{figure}

\begin{figure}
    \centering
  \includegraphics[width=\linewidth, keepaspectratio]{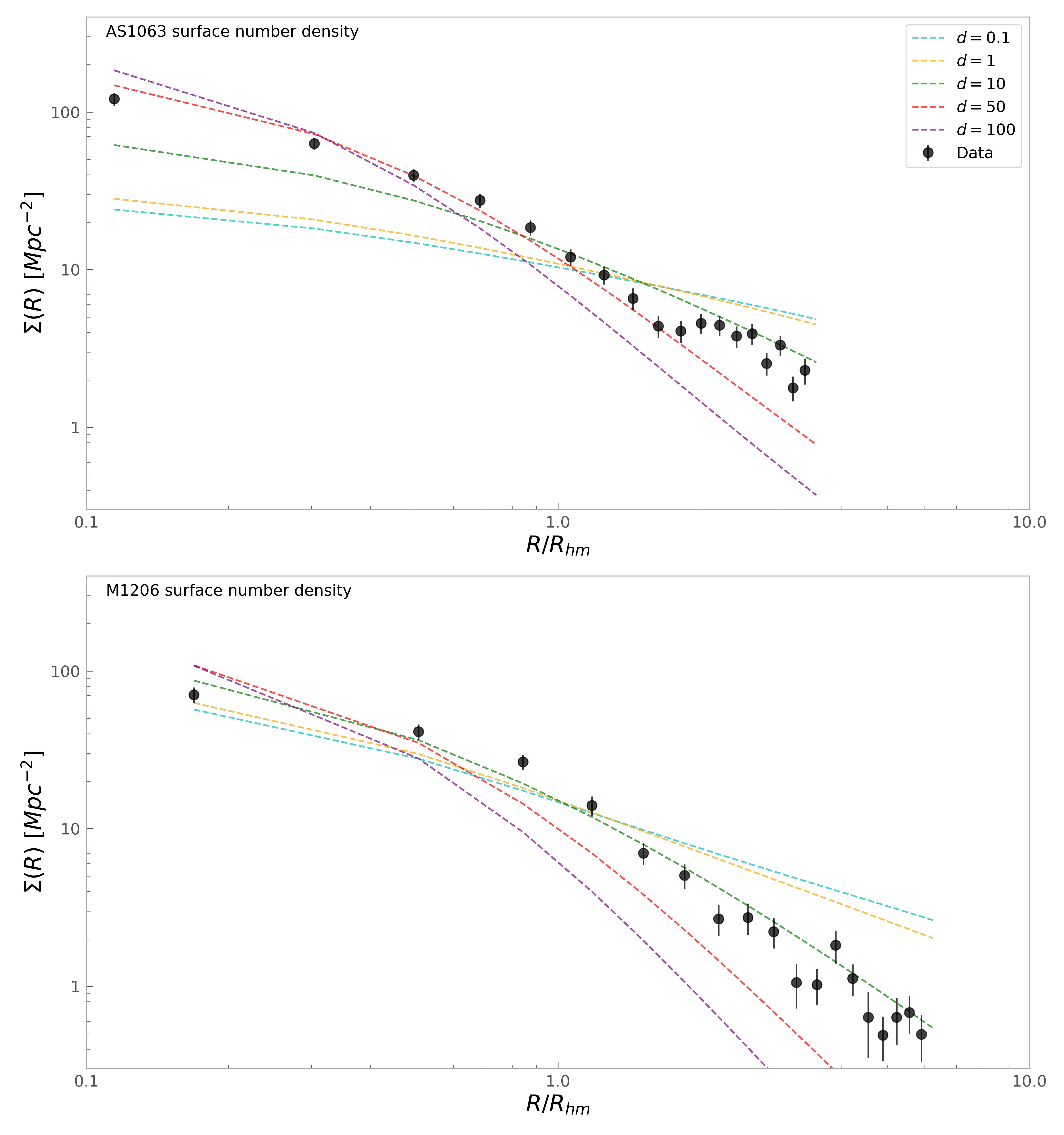}
  \caption{The projected number density profile of the galactic component derived from the $f^{(\nu)}$ distribution function with fitted parameters for AS1063 and M1206 (see Table \ref{table:summary_cap_6_5}). 
  The plot shows in black the data points derived from photometric data (see Fig. \ref{fig:radial_surface_number_density}), superimposed on the projected number density profile derived from the moments of the distribution function, for 5 choices of the value of the parameter $d$ (see Fig. \ref{fig:DF_beta} for the associated anisotropy profile). 
  The value of the parameter $A$ is obtained after a best fit against the observed profile performed within $2 R_\text{hm}$.
  Both clusters are well described by the curves associated to a value of $d\approx 10$-$50$, confirming the results of Fig. \ref{fig:DF_vel_dispersion}.}
  \label{fig:F_nu_COMP_4sept_surface_density_FIN}
\end{figure}

\begin{figure}
    \centering
  \includegraphics[width=\linewidth, keepaspectratio]{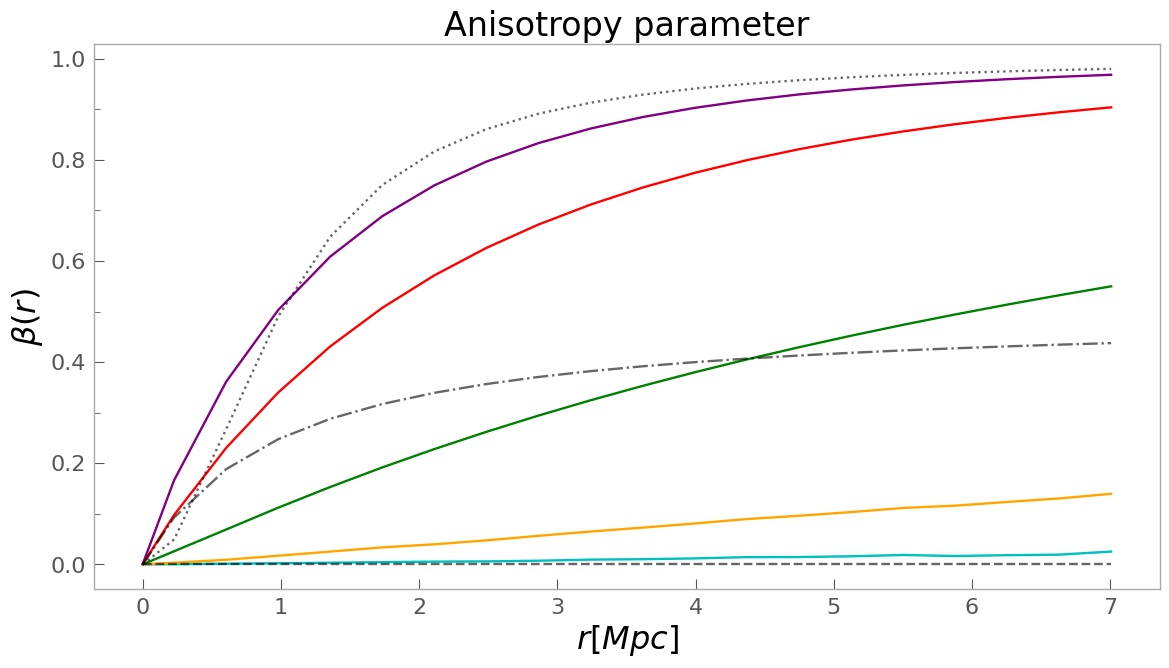}
  \caption{The colored solid lines show the anisotropy profile as a function of the free parameter $d$. The distribution of stellar orbits associated with the adopted distribution function is characterized by isotropy in the central region and radially biased anisotropy in the outer parts, which becomes more prominent as the value of $d$ increases. The dashed, dashed-dotted and dotted gray curves represent, respectively, the isotropic, radial and Osipkov-Merritt anisotropy profiles of Table \ref{tab:beta_param}.}
  \label{fig:DF_beta}
\end{figure}

\indent The distribution function that we will consider is (see \citealt{F_nu_stiavelli}; with $\nu = 1$)
\begin{equation}\label{eq:f_nu}
    f^{(\nu)}= A \exp\left( -a E - d\frac{J}{|E|^{3/4}} \right),
\end{equation}
where the specific energy $E$ and the specific angular momentum $J$ are defined as $E = v^2/2 + \Phi(r)$ and $J= |\bf{r} \times \bf{v}|$.\\
\indent In the fully self-consistent case, in which the ``particles” described by the distribution function determine the mean gravitational potential, the boundary conditions to the Poisson equation impose a relation between the depth of the potential well $\Phi(0)$ and a dimensionless parameter that depends on $A$, $a$, and $d$. Here the mean potential is external and imposed \REV{(in our model, we refer to the total mass profile described by \citealt{Umetsu_strong_weak_lens_CLASH} and introduced in Sect. 6.2)}. Therefore, the three (positive) parameters can be adjusted independently to produce a best-fit to the available data for the galactic component assumed to be described by the above distribution function. The parameter $A$ sets the scale of the density profile,  whereas the parameter $a$ sets the scale of the velocity dispersion profile. The third parameter $d$ can be adjusted to determine the radius beyond which the velocity dispersion is significantly anisotropic. \\
\indent The relevant calculations are best carried out after a suitable reduction to dimensionless variables. In particular, we define the imposed dimensionless potential as $\psi = - a \Phi$ and refer to the dimensionless radial coordinate $x = r/r_s$, where for each cluster $r_s$ is the scale of the best-fit NFW potential of Table \ref{tab:umetsu_nfw_totalmass}, so that $\psi(x) = \Psi x^{-1}\ln{(1+x)}$. By inserting the imposed potential in Eq. (\ref{eq:f_nu}), from the distribution function it is possible to compute the volume density $n(r)$, the radial velocity dispersion $\sigma_{rr}(r)$, and the local anisotropy profile $\beta(r$). By suitable integration, these profiles can be converted into projected profiles $\Sigma(R)$ and $\sigma_{los}(R)$ and compared with the observations. For each cluster, this procedure leads to the determination of the best-fit parameters $A$, $a$.\\
\indent \REV{In Figs. \ref{fig:DF_vel_dispersion}-\ref{fig:F_nu_COMP_4sept_surface_density_FIN} we see the how the line-of-sight velocity dispersion parameter and projected number density profiles vary as a function of $d$.
We note that the parameter range $d=10-50$ leads to a self-consistent dynamical model that matches well the photometric and spectroscopic observations, and that is characterized by isotropic pressure in the center and radial pressure anisotropy in the outer regions (see Fig. \ref{fig:DF_beta}) as also indicated by the results based on the Jeans equations in the previous Subsection.}\\
\indent The velocity dispersion anisotropy profiles with $d=10-50$ suggest the presence of mild anisotropy in the core $\beta(0.5')\approx0.05$. 
If the measurements of the anisotropy profiles are accurate in the very central regions, we may argue that both cores do not follow the standard $(V/\sigma, \varepsilon)$ relation mentioned in Sect. 4.1 .\\
\indent For a detailed description, see Appendix A. In Appendix B, we test the robustness of our sample by imposing a cut in magnitude at $m_{R_c}\leq 22$.

\begin{table}
    \centering
        \begin{tabular}{ ccccc }
         \hline \hline
               & $r_s$        & $\Psi$ & $   A$                         & $d$                       \\
               &       [Mpc]  &        &   [s$^3$ km$^{-3}$ Mpc$^{-3}$] & [km$^{-1/2}$ s$^{-1/2}$]  \\
         \hline 
         AS1063& $0.66$       & 5.1    & $0.5-3.3\times10^{-10}$        & $10-50$                   \\
         \hline
         M1206 & $0.60$       & 7.7    & $1.0-4.1\times10^{-11}$        & $10-50$                   \\
         \hline \hline
        \end{tabular}
    \caption{Distribution function parameters. The scale radius of the external potential $r_s$ is taken from \cite{Umetsu_strong_weak_lens_CLASH}, and the concentration parameter $\Psi$ depends on the parameters used to describe the NFW potential and $a$ as described in Sect. 6.3. The values of $A$ and $d$ are those associated with a reasonable fit to the available data.}
    \label{table:summary_cap_6_5}
\end{table}

\section{Conclusions}

In this paper we have investigated some dynamical properties of the galactic component of two clusters of galaxies: Abell S1063 and MACS J1206.
In particular, we have addressed the question of the possible presence of systematic rotation and the issue of their general relaxation state.
We find evidence for differential rotation in both clusters, with slow rotation ($V_\text{r}/\sigma \approx 0.15$) in the centers tapering off at larger radii, and pressure anisotropy.
Differential rotation and pressure anisotropy can contribute to produce nontrivial gradients in the phase space. 
Therefore, it would be useful to extend this study to more complex dynamical models (for an example developed for globular clusters, see \citealt{Varri_Bertin}).
It would be interesting to check whether the ICM component of these two clusters is also characterized by lack of systematic rotation; an X-ray measurement of this property is likely to become available in the near future 
(see \citealt{ICM_rotation}, \citealt{Liu_Tozzi_2019}).

As to the relaxation state of the two clusters, no signs are found of energy equipartition, but there is a clear indication of (stellar) mass segregation. The presence of mass segregation in the absence of equipartition suggests that the observed mass segregation is associated with mechanisms that are unrelated to the possible role of collisional processes (galaxy-galaxy encounters). 

Indeed, velocity dispersion anisotropy is found to be present in the two clusters and turns out to be qualitatively similar to that found in violently relaxed collisionless systems (see Subsect. 6.2). This last conclusion is strengthened by the overall success in matching the observations with the predictions of a physically justified distribution function with an imposed external potential, as described in Subsect. 6.3.

\section*{Acknowledgments}
\REV{We thank the anonymous referee for useful comments that improved the presentation.}
We acknowledge financial support through grants PRIN-MIUR 2017WSCC32 and 2020SKSTHZ, and INAF “main-stream” 1.05.01.86.20 and 1.05.01.86.31.

\bibliographystyle{aa}
\bibliography{references}

\begin{appendix}
\section{Some properties of the distribution function}

The choice of a distribution function (DF) completely describes the dynamics of a system, and its integration gives self-consistent photometric and kinematic profiles (e.g. see \citealt{Bertin_Dynamics_gal_2nd}).

Here we discuss some of the properties of the DF that we adopted to describe the galactic component.
The $f^{(\nu)}$ DF was originally conceived to incorporate the picture of formation by incomplete violent relaxation (in particular, see \citealt{F_nu_stiavelli}, \citealt{Bertin_Stivelli}, \citealt{Trenti_bertin}).
The family of distribution functions is derived under the assumption that only one additional quantity, $Q=J^{\nu}(-E)^\frac{3}{4\nu}$, is conserved in addition to the total energy and the number of particles. 
With this assumptions and limiting the calculation to the spherical symmetric case, the extremization of the entropy leads to the expression of $f^{(\nu)}$ (with $\nu = 1$) as
\begin{equation}\label{eq:f_nu_appdx}
    f^{(\nu)}= A \exp\left( -a E - d\frac{J}{|E|^{3/4}} \right),
\end{equation}
where the specific energy $E$ and the specific angular momentum $J$ are defined as $E = v^2/2 + \Phi(r)$ and $J= |\bf{r} \times \bf{v}|$.
Dimensionally we see that $A=[M L^{-6} T^3]$, $a=[L^{-2} T^2]$ and $d=[L^{-1/2} T^{-1/2}]$.

Given a positive-definite distribution $g(x)$ defined on an interval $[a,b]$, one defines the $n^\text{th}$-moment of such distribution as
\begin{equation}\label{eq:df_moment}
\langle x^n \rangle = \int_a^b g(x) x^n \dd{x} 
\end{equation}
Accordingly, the density $\rho(r)$ is the zeroth moment of $f(r, \vec{v},t)$ integrating over the velocities
\begin{equation}\label{eq:df_density}
\rho(r, t) = \int f(r, \vec{v}, t) \dd^3 v
\end{equation}
and the velocity dispersion is the second moment weighted over the density
\begin{equation}\label{eq:vel_disp_DF_appdx}
\sigma^2(r) = \frac{\int v^2 f(r,\vec{v}) \dd^3v}{\int f(r,\vec{v}) \dd^3v} \text{ .}
\end{equation}

As discussed in Sect. 6, the three (positive) parameters $A, a, d$ can be adjusted independently to produce a best-fit to the available data for the galactic component assumed to be described by the above distribution function. 
The parameter $A$ sets the scale of the density profile, whereas the parameter $a$ sets the scale of the velocity dispersion profile.
The third parameter $d$ can be adjusted to determine the radius beyond which the velocity dispersion is significantly anisotropic.

The relevant calculations are best carried out after a suitable reduction to dimensionless variables. 
In particular, we define the imposed dimensionless potential as $\psi(r) = - a \Phi(r)$ and refer to the dimensionless radial coordinate $x = r/r_s$, where for each cluster $r_s$ is the scale of the best-fit NFW potential of Table \ref{tab:umetsu_nfw_totalmass}, so that $\psi(x) = \Psi x^{-1}\ln{(1+x)}$. By inserting the imposed potential in Eq. (\ref{eq:f_nu}), from the distribution function it is possible to compute the volume density $n(r)$, the radial velocity dispersion $\sigma_{rr}(r)$, and the local anisotropy profile $\beta(r$). By suitable integration, these profiles can be converted into projected profiles $\Sigma(R)$ and $\sigma_\text{los}(R)$ and compared with the observations.

To obtain an estimate for $a$, we can evaluate Eq. (\ref{eq:vel_disp_DF_appdx}) for $r=0$, and after solving the Gaussian integral we obtain
\begin{equation}\label{eq:central_vel_disp_a_DF_appdx}
    \sigma^2(0) = \frac{2}{a}\left[\frac{3/2\sqrt{\pi}\erf{(\sqrt{\Psi})}-(2\Psi^{3/2}+3\sqrt{\Psi})e^{-\Psi}}{\sqrt{\pi} \erf{(\sqrt{\Psi})}-2\sqrt{\Psi}e^{-\Psi}} \right] \approx \frac{3}{a} \text{ .}
\end{equation}

We then can fit this value with the measured central velocity dispersion and obtain an estimate for the parameter $a$.
Here we are ignoring the projection effect and the fact that the innermost observation are not at the exact origin $r=0$. 
We can use this first estimate as a prior to then fit the value of the parameter $a$ with the central projected line-of-sight velocity dispersion (wich depends weakly from the value of $d$)
\begin{equation}\label{eq:central_vel_disp_proj_a_DF_appdx}
\begin{split}
    &\sigma^2_\text{los}(0) = \\
    &\frac{2}{a}\frac{\int_0^\infty \int_0^\pi \int_0^{\sqrt(\psi)}\exp\left[ -w^2 +\psi-\frac{\sqrt(2)\xi w \sin\vartheta}{|w^2-\psi|^{3/4}}\right]w^4\cos^2\vartheta\sin\vartheta \dd w \dd \vartheta \dd \xi}
    {\int_0^\infty \int_0^\pi \int_0^{\sqrt(\psi)}\exp\left[ -w^2 +\psi-\frac{\sqrt(2)\xi w \sin\vartheta}{|w^2-\psi|^{3/4}}\right]w^2\sin\vartheta \dd w \dd \vartheta \dd \xi}
\end{split}
\end{equation}
after an appropriate dimensionless scaling of the quantities $\xi = (a^{1/4}d)r$, $w^2 = (a/2)v^2$. 

From Eq. (\ref{eq:df_density}) we obtain
\begin{equation}
\begin{split}
    &\rho(\xi)  =  \\
    &A\left(\frac{2}{a}\right)^{3/2}\int_0^\pi \int_0^{\sqrt{\psi}} \exp\left[  -w^2 +\psi -\frac{\sqrt{2}\xi w \sin{\vartheta}}{|w^2-\psi|^{3/4}} \right]w^2 \sin{\vartheta} \dd w \dd \vartheta
\end{split}
\end{equation}
The value of the parameter $A$ can be found fitting the surface number density after projecting Eq. (\ref{eq:df_density}) through Eq. (\ref{eq:surface_projection_cap_6}). 

At last, we can derive the anisotropy parameter profile $\beta$

\begin{equation}\label{eq:beta_DF_appdx}
\begin{split}
    \beta(\xi) = & 1 - \frac{\sigma_T^2}{2\sigma_r^2} = \\
    1 - &\frac{1}{2}\frac{\int_0^\pi \int_0^{\sqrt{\psi}} \exp\left[  -w^2 +\psi -\frac{\sqrt{2}\xi w \sin{\vartheta}}{|w^2-\psi|^{3/4}} \right]w^4 \sin^3{\vartheta} \dd w \dd \vartheta}
    {\int_0^\pi \int_0^{\sqrt{\psi}} \exp\left[  -w^2 +\psi -\frac{\sqrt{2}\xi w \sin{\vartheta}}{|w^2-\psi|^{3/4}} \right]w^4 \cos^2{\vartheta}\sin{\vartheta} \dd w \dd \vartheta} \text{ ,}
\end{split}
\end{equation} 

where the tangential component of the velocity dispersion tensor is $\sigma^2_T = \sigma_{\theta\theta}^2 + \sigma_{\phi\phi}^2$ and the radial component $\sigma^2_{rr}
$.
After fixing the value of $a$, the profile $\beta(r)$ depends only on the choice of the value of the parameter $d$.

Noting that $\beta(0)=0$, the first order MacLaurin expansion of $\beta(r)$ is
\begin{equation}\label{eq:beta_first_order_appdx}
    \beta(r) \sim r \frac{9\pi a^{1/4}d}{16\sqrt{2}}\left[\int_0^{\sqrt{\Psi}}e^{-w^2}w^4\dd w\right]^{-1}\int_0^{\sqrt{\Psi}}\frac{e^{-w^2}w^5 \dd w}{|w^2-\Psi|^{3/4}}\text{ ,}
\end{equation}
which indicates that the isotropic orbits in the center evolve towards radially biased orbits for ($r>0$).

\section{Sample robustness}

To test the dependence of our conclusions on the radial and brightness completeness of our sample, we selected a smaller sample for both clusters excluding the faintest galaxy members by imposing a cut in magnitude at $m_{R_c}\leq 22$. This yielded 648 galaxy members for AS1063 and 310 for M1206.
We ran the same analysis performed on the full sample on this smaller sample, and we found that the velocity dispersion and half-mass radius vary by $\leq 5\%$. 
Furthermore, the phase-space distribution of the smaller sample gives us results fully consistent with the results in Sections 4, 5, and 6.

\section{Rotation signatures at intermediate and large radii}
\begin{figure*}
  \centering
  \includegraphics[width=\linewidth, keepaspectratio]{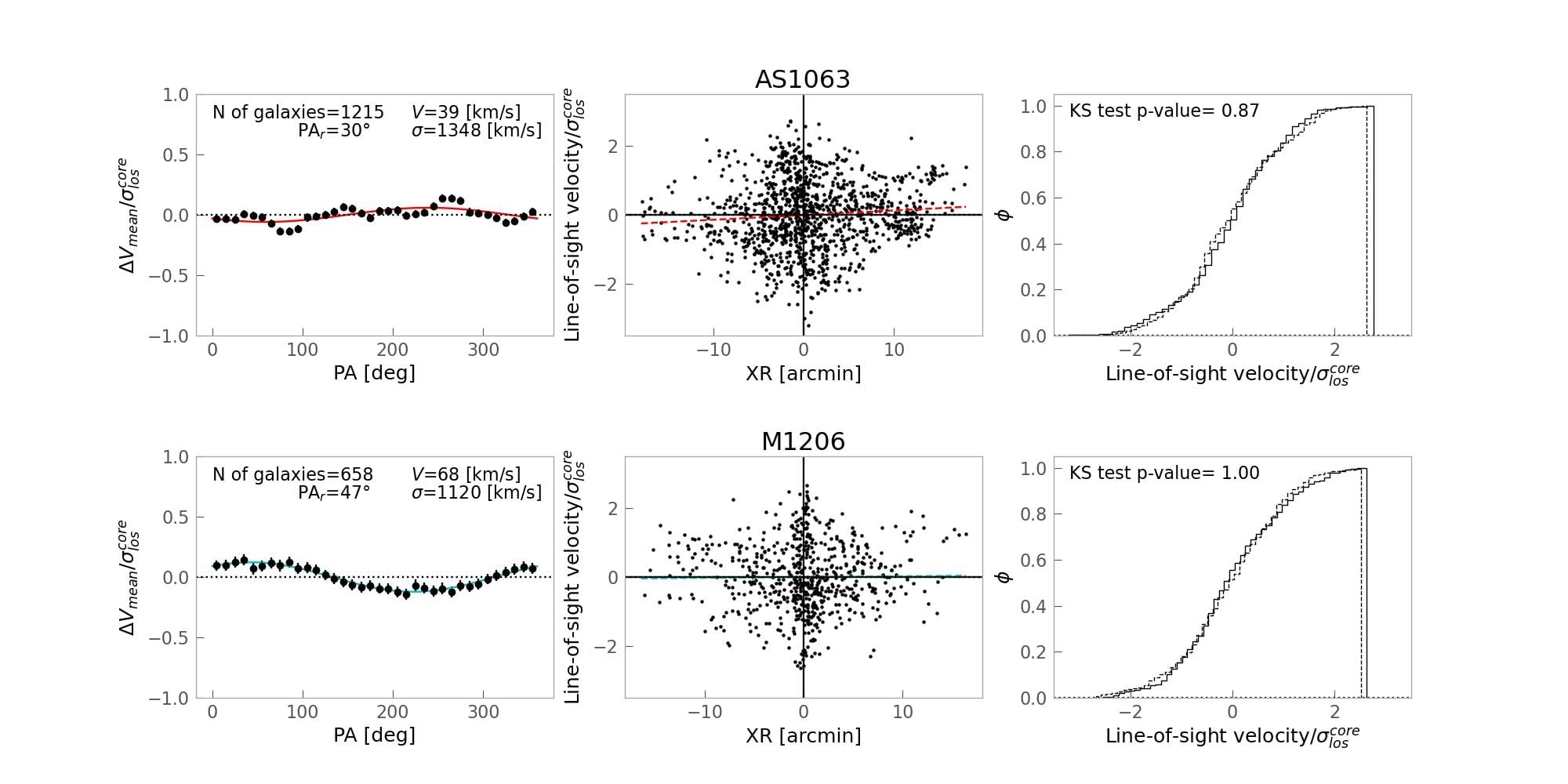}
  \caption{
  Diagnostic diagrams of the rotation signature detected in the clusters, considering all their galaxy members. See Fig. \ref{fig:central_rotation}.}
  \label{fig:all_GC_rotation}
\end{figure*}

\begin{figure*}
  \centering
  \includegraphics[width=\linewidth, keepaspectratio]{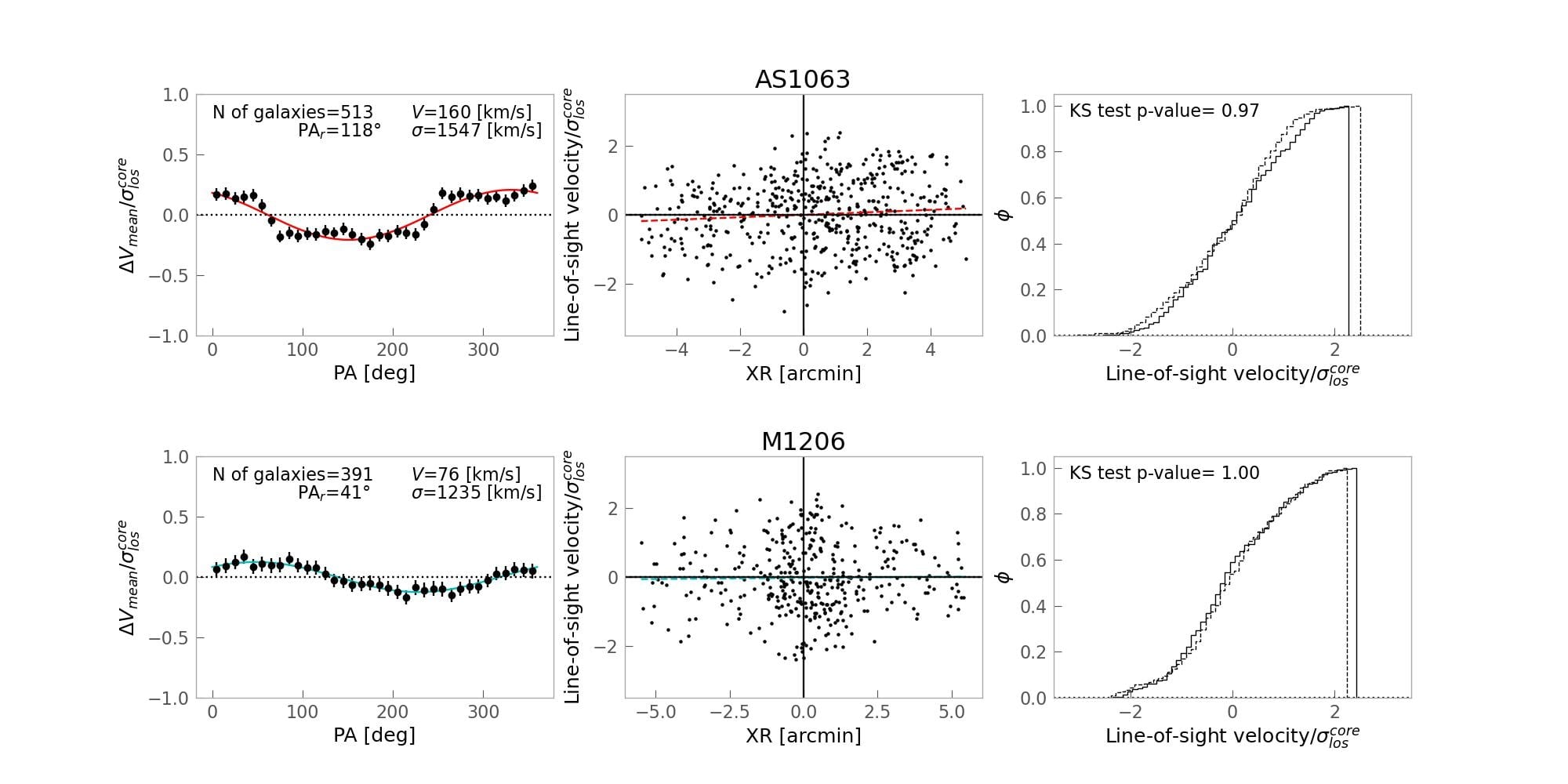}
  \caption{
  Diagnostic diagrams of the rotation signature detected in the clusters within $5'$ from the center. See Fig. \ref{fig:central_rotation}.}
  \label{fig:5arcmin_GC_rotation}
\end{figure*}

Figs. \ref{fig:all_GC_rotation}-\ref{fig:5arcmin_GC_rotation} represent the results of the analysis introduced in Sect. 4.1 (see also Fig. \ref{fig:central_rotation}) on clusters as a whole, and within $5'$ from the centers.
Fig. \ref{fig:all_GC_rotation} confirms that there is no evidence of rotation when considering the complete samples, while Fig. \ref{fig:5arcmin_GC_rotation} shows a slight increase in $\Delta V_\text{mean}/\sigma_\text{los}$ as expected from the results discussed in Sect. 4.1.

\FloatBarrier
\FloatBarrier

\section{Jeans equations with NFW density}
\begin{figure}[h]
  \centering
  \includegraphics[width=\linewidth, keepaspectratio]{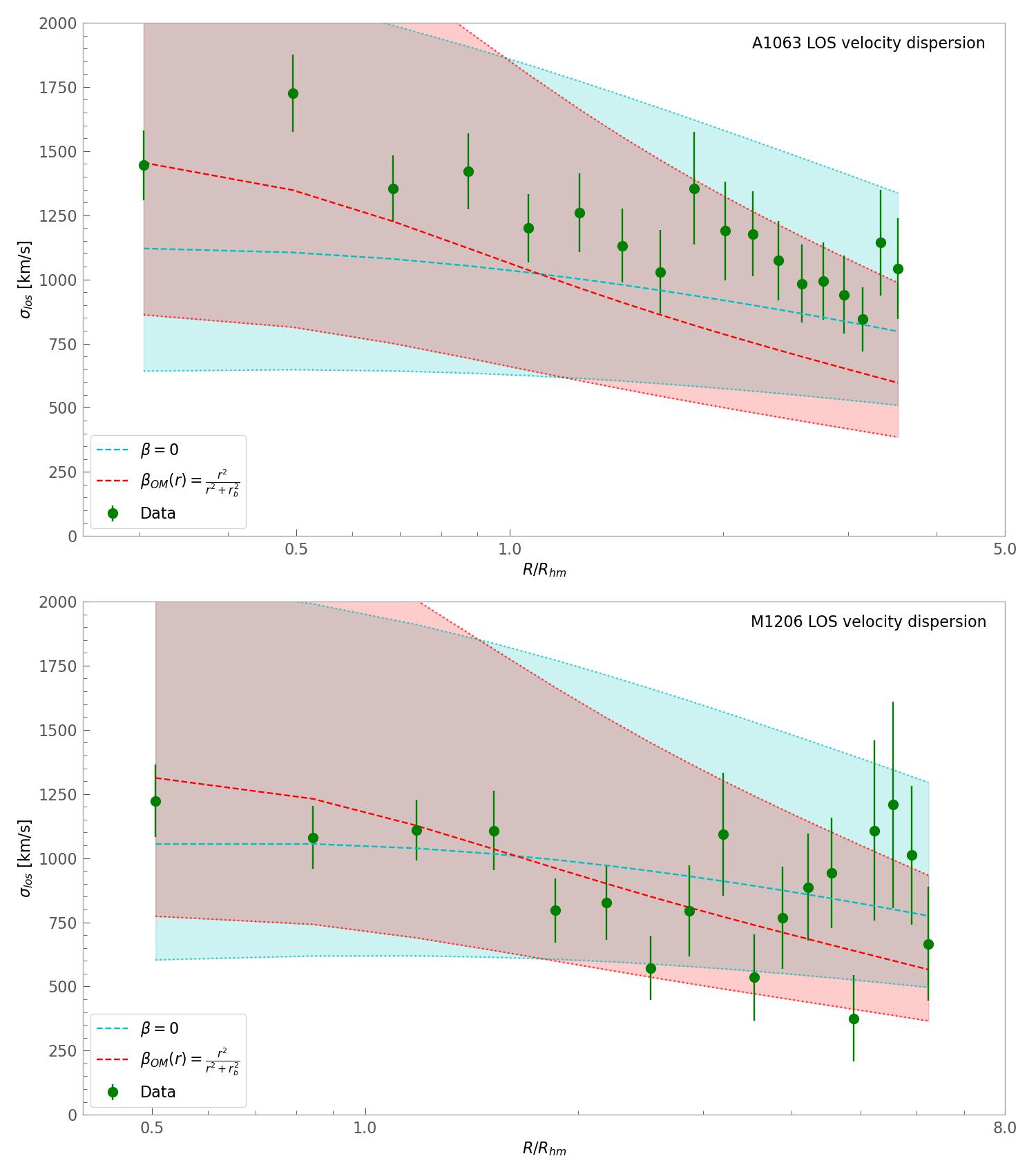}
  \caption{\REV{The line-of-sight velocity dispersion profile of the galactic component obtained from the Jeans equation for AS1063 and M1206, using a NFW profile for the galaxy number density. See Fig. \ref{fig:Jeans_vel_disp}.}}

  \label{fig:Jeans_vel_disp_NFW}
\end{figure}

\REV{Fig. \ref{fig:Jeans_vel_disp_NFW} shows the result for the line-of-sight obtained from the Jeans equation, adopting the NFW profile obtained in Subsect. 6.1.}

\FloatBarrier
\end{appendix}

\end{document}